\documentclass[a4paper,11pt]{article}
\pdfoutput=1 

\usepackage{jcappub} 

\usepackage[T1]{fontenc} 
\usepackage{bm}
\usepackage{extarrows}

\def\diff{{\rm d}}

\newcommand{\e}[0]{{\rm e}}

\def\bgam{\bm\gamma}

\def\Plin{P_{\rm Lin}}
\def\DIII{\delta^{\rm D}_{\rm 3D}}
\def\DII{\delta^{\rm D}_{\rm 2D}}
\def\DI{\delta^{\rm D}_{\rm 1D}}
\def\deltaK{\delta^{\rm K}}

\newcommand\Eqn[1]     {Eq.\,(\ref{#1})}
\newcommand\Eqns[2]    {Eqs\,(\ref{#1}) and~(\ref{#2})}

\newcommand\Figure[1]     {Figure\,{\ref{#1}}}
\newcommand\Fig[1]     {Fig.\,{\ref{#1}}}

\newcommand\nn         {\nonumber}

\newcommand{\be}{\begin{equation}}
\newcommand{\ee}{\end{equation}}

\newcommand{\ba}{\begin{eqnarray}}
\newcommand{\ea}{\end{eqnarray}}

\newcommand{\bes}{\begin{subequations}}
\newcommand{\ees}{\end{subequations}}

\def\ex{{\rm e}}

\def\omr{\Omega_{\rm r}}
\def\oml{\Omega_{\Lambda}}
\def\omk{\Omega_{k}}

\def\omm0{\Omega_{\rm m,0}}
\def\omr0{\Omega_{\rm r,0}}
\def\oml0{\Omega_{\Lambda,0}}
\def\omk0{\Omega_{k,0}}

\def\Msol{h^{-1}M_{\odot}}
\def\kpc{\, h^{-1}{\rm kpc}}
\def\Mpc{\, h^{-1}{\rm Mpc}}

\def\kMpc{\, h \, {\rm Mpc}^{-1}}

\def\FLC{F^{\rm LC}}
\def\DLC{\Delta\!^{\rm LC}}

\def\1D{{\rm 1D}}
\def\2D{{\rm 2D}}
\def\3D{{\rm 3D}}

\def\Vmu{{V^{\rm max}_{\mu}}}

\def\Vu{{V_{\mu}}}

\def\g{{\rm g}}

\def\br{{\bf r}}
\def\bx{{\bf x}}

\def\bk{{\bf k}}
\def\bq{{\bf q}}

\def\dr{{\rm d}^3{\bf r}}

\def\dk{{\rm d}^3{\bf k}}

\def\1Loop{{\rm 1Loop}}

\def\Fg{{\mathcal F}}
\def\Lmin{{L_{\rm min}}}

\title{The galaxy power spectrum on the lightcone: deep, wide-angle redshift surveys and the turnover scale}

\author[a]{Dan Pryer,}
\author[a]{Robert E. Smith,}
\author[a]{Robin Booth,}
\author[b]{Chris Blake,}
\author[c,d]{Alexander Eggemeier}
\author[a]{and Jon Loveday}

\affiliation[a]{Astronomy Centre, Department of Physics \& Astronomy,
        University of Sussex,\\ Brighton, BN1 9RH, UK}
\affiliation[b]{Centre for Astrophysics \& Supercomputing,
  Swinburne University of Technology,\\ P.O. Box 218, Hawthorn, VIC 3122, Australia}
\affiliation[c]{Institute for Computational Cosmology, Department of Physics,
        Durham University,\\ South Road, Durham DH1 3LE, United Kingdom}
\affiliation[d]{Argelander-Institut für Astronomie,
  Auf dem Hügel 71, D-53121 Bonn, Germany}

\emailAdd{d.pryer@sussex.ac.uk}
\emailAdd{r.e.smith@sussex.ac.uk}

\abstract{We derive expressions for the survey-window convolved galaxy power
spectrum in real space for a full sky and deep redshift survey, but
taking into account the geometrical lightcone effect. We investigate
the impact of using the standard mean redshift approximation as a
function of survey depth, and show that this assumption can lead to
both an overall amplitude suppression and scale-dependent error when
compared to the `true' spectrum. However, we also show that by using a
carefully chosen `effective fixed-time', one can find a range of
scales where the approximation to the full model is highly accurate,
but only on a more restricted set of scales. We validate the theory by
constructing dark matter and galaxy lightcone mock surveys from a
large $N$-body simulation with a high cadence of snapshots. We do this
by solving the light cone equation exactly for every particle, where
the particle worldlines are obtained in a piecewise fashion with cubic
interpolation between neighbouring snapshots. We find excellent
agreement between our measurements and the theory ($\sim \pm 5\%$)
over scales $(0.004\kMpc \leq k \leq 0.54\kMpc)$ and for a variety of
magnitude limits. Finally, we look to see how accurately we can measure the
turnover scale of the galaxy power spectrum $k_0$. Using the lightcone
mocks we show that one can detect the turnover scale with a
probability $P \geq 95\%$ in an all-sky catalogue limited to an
apparent magnitude $m_{\rm lim}\sim 21$. We also show that the
detection significance would remain high for surveys with $m_{\rm
  lim}\sim22$ and $20\%$ sky coverage.}

\begin{document}
\maketitle
\flushbottom


\section{Introduction}
\label{sec:LC1:intro}

The next decade will herald significant advances in the quantity of
high fidelity data for mapping the large-scale distribution of
galaxies. This will be facilitated through a series of exciting new
wide-field observatories, such as the spectroscopic missions
\href{https://www.desi.lbl.gov}{DESI}\footnote{www.desi.lbl.gov}
\cite{DESI2016},
\href{http://www.4most.eu/cms/}{4MOST}\footnote{www.4most.eu/cms/}
\cite{4MOSTCRS2019}, the \href{https://sci.esa.int/web/euclid}{ESA
  Euclid}\footnote{sci.esa.int/web/euclid} space mission
\cite{Euclid2011} and the NASA mission {\tt SPHEREx}
\cite{Doreetal2014}. These will be complemented by the panchromatic
imaging surveys like
\href{https://www.lsst.org}{LSST}\footnote{www.lsst.org}
\cite{LSST2009} and Euclid. Together these observatories will map the
galaxy distribution out to high redshift and over significant
fractions of the full-sky, and will thus chart effective volumes that
approach the observable size of our Universe out to a given redshift
\cite{Doreetal2014}. All of these surveys will be able to make
unprecedented measurements of the statistical properties of the
spatial distribution of galaxies.

The lowest order spatial statistic of interest is the correlation
function, or equivalently its Fourier-space dual the power spectrum.
The full shape and amplitude of the matter power spectrum carries a
great deal of information about the cosmological model and the
primordial distribution of the initial density fluctuations
\cite{Tegmark1997}.  However, the extraction of this information is
complicated by a number of effects: first, the initial linear density
perturbations become nonlinearly coupled due to gravitational
instability \cite{Peebles1980,Bernardeauetal2002}. Understanding this
process is further complicated by the fact that the matter
distribution is built from the weighted densities of baryons, cold
dark matter and massive neutrinos, each component of which needs to be
propagated into the nonlinear regime through careful modelling
\cite[see for
  example][]{Jingetal2006,SomogyiSmith2010,McCarthyetal2018}. Second,
galaxies are a biased set of tracers for the matter distribution, and
may in full generality be a nonlinear, time-dependent, non-local,
stochastic sampling of the matter, and this connection varies with the
types of galaxies that are selected
\cite{Kaiser1984,FryGaztanaga1993,Coles1993,DekelLahav1999,
  Baldaufetal2012,Chanetal2012,Eggemeieretal2021}. Third, owing to the
fact that galaxy positions are inferred from their measured redshifts
there is a distortion in the mapping from observed to true position,
which arises due to the fact that galaxies have peculiar velocities
relative to the Hubble flow -- redshift space distortions
\cite{Kaiser1987,Hamilton1998,Scoccimarro2004}.

To this one can also add the lightcone effect, which has surfaced in
this recent era of precision cosmology -- for deep surveys one has to
account for the fact that we are correlating galaxy positions on the
observer's past lightcone
\cite{Yamamotoetal1999,Matsubara2000}. Furthermore, there are
additional corrections that arise due to general relativistic effects
and also the magnification effects that all arise due to light
propagation in a weakly perturbed spacetime
\cite{YooSeljak2015,Bertaccaetal2018}. Another source of error arises
from the assumption of a wrong cosmological model, in principle this
could be overcome by simply recalculating the statistic of interest
for every assumed cosmological model to be tested
\cite{AlcockPaczynski1979,Ballingeretal1996}.  Together, these effects
can all be thought of as modelling problems and one should look to
account for them when comparing to measurements.  Besides these
`physical' effects one can add a number of important `observational'
effects that also need to be carefully taken into account: the survey
mask; star--galaxy classification; curved sky effects; fibre collision
and completeness; accurate flux calibration across the survey;
K-corrections, etc.

In this paper we focus on a subset of these issues. Our goal here is
to build up the methodology for forward modelling the theory of the
real space, 3D galaxy power spectrum in a survey which spans a
significant fraction of the sky and thus requires us to take into
account wide-angle clustering effects and one that is also deep and so
requires us to handle the lightcone effect. This means that we will
need to take into account: the convolution with the survey window
function, the lightcone effect, including the unequal time correlators
and work on the curved sky. We reserve exploration of redshift-space
distortions and the power-spectrum multipoles for future work.  Our
aim here is thus to evolve the estimation and forward modelling
analysis methodology to the next level
\cite{Feldmanetal1994,Tegmark1997b,Yamamoto2003,Percivaletal2004a,SmithMarian2015}.

Regarding the lightcone effect, our work here is most closely inspired
by that of \cite{Yamamotoetal1999} and we provide alternate
derivations of results from that work that are easier to evaluate, but
our results go beyond that work in a number of key ways. First,
working with the spherical-Fourier-Bessel (hereafter sFB) expansion of
the theory, we find expressions that are simpler to compute. Second,
we include galaxy weighting and luminosity dependent galaxy
bias. Third, we derive the shot-noise and show that its form remains
as that found by \citep{Percivaletal2002}. Fourth, we include general
expressions with a survey mask. We validate all of this with a series
of deep full-sky galaxy mocks derived from $N$-body simulations.

In this work we will make use of the sFB harmonic analysis of a galaxy
redshift survey and this approach has a long history in cosmology
\cite{Peebles1973, Scharfetal1992, ScharfLahav1993, Fisheretal1995,
  HeavensTaylor1995, Tadrosetal1999, Percivaletal2004b,
  Nicolaetal2014, Lanusseetal2015, Shiraishietal2020}, and we will
take many of our cues from these important works.  Furthermore, as
discussed in \citep{CastorinaWhite2018}, for a more exhaustive
analysis of galaxy clustering for wide-angled surveys two schools of
thought have emerged in recent years: one is to develop the power
spectrum multipoles, following the estimator style of
\cite{Yamamotoetal2006,Scoccimarro2015}; the other is to work in the
sFB framework from the start. The former approach has the problem of
requiring complex estimators, but is easier to model. The latter
allows us to make use of the decomposition of the problem into radial
and angular modes, but has been highlighted as harder to model. For
our work, since we are restricting our attention to the monopole of
the power spectrum, there is no great complexity in how we estimate it
\cite{Scoccimarro2015} from data.  That is not to say that there are
no subtleties in its estimation. Our effort therefore mainly goes into
modelling this signal for a wide-angle survey, that is also deep.

An additional important science driver for future surveys is the
question of: How well can we detect the `turnover scale' of the matter
power spectrum. This length scale denotes the point of maximum
amplitude of the matter power spectrum and for the $\Lambda$CDM model
is a direct imprint of the epoch of matter-radiation equality in the
early stages of the Universe. A number of previous works have examined
this issue
\cite{BlakeGlazebrook2003,SeoEisenstein2003,BlakeBridle2005}. However,
here we now relax some of the simplifying assumptions and use our
improved formalism to attempt to address questions on this topic.

The paper breaks down as follows: In \S\ref{sec:LC1:theory} and
\S\ref{sec:LC1:LCtheory} we develop the theory of the galaxy power
spectrum on the curved-sky, past lightcone. In \S\ref{sec:approx} we explore 
some approximate forms to our model that can make it more numerically
tractable. Then in \S\ref{sec:LC1:mocks} we describe how we create mock lightcone
catalogues, detail the process used to estimate the power spectrum
from these mocks, and then make comparisons to theory in \S\ref{sec:LC1:results}. 
We then explore the ability to constrain the 
power spectrum turnover scale from our measured spectra in \S\ref{sec:turnover}.
Finally, in \S\ref{sec:LC1:conclusion} we
summarise our findings, conclude and discuss future work.


\section{Theoretical background}\label{sec:LC1:theory}

We now review some of the key aspects of the background theory
necessary for what follows. Note that we will neglect the effects of
time delay and photon deviations that are required to describe a full
general relativistic treatment of light propagation on the perturbed
past lightcone.


\subsection{The past lightcone}

To begin, we assume that the unperturbed background spacetime is a
spatially flat, homogeneous and isotropic, spacetime. Hence, in
spherical coordinates the differential FLRW line element can be
written as:
\be ds^2 = a(\eta)^2\left[c^2d\eta^2-d\chi^2-
  \chi^2 d\theta^2-\chi^2 \sin^2\theta d\phi^2\right] \ ,\ee
where $\eta$ is the conformal time, $a(\eta)$ is the scale factor,
$\chi$ is the comoving radial geodesic distance, $c$ is the speed of
light and $\theta$ and $\phi$ are the usual polar angles, which we
denote collectively as $\bgam\rightarrow (\theta,\phi)$. A light signal emitted
at spacetime point $(\eta_1,\chi_1,\bgam_1)$ will be received by an
observer located at the origin of our coordinate system at spacetime
point $(\eta_2,0,{\bf 0})$. From the FLRW line element, we see that
points on the past lightcone are required to obey the following
relation between the radial comoving distance from the observer and
the redshift:
\be r(z) = \left| \int^0_{r} d\chi \right| =
\int_{\eta_1}^{\eta_2} c d\eta = \int_{0}^{z}\frac{c
  dz'}{H(z')}\ , \label{eq:chi}\ee
where the redshift is related to the expansion factor as
$z=1/a(\eta)-1$ and the Hubble rate is $H(z)\equiv a^{-2}(da/d\eta)$. This
can be calculated from the Friedmann equations, and for the case of
flat $\Lambda$CDM is given by:
\be
H^2(z) = H_0^2\left[\Omega_{\rm m,0}(1+z)^{3}+\Omega_{\Lambda,0}\right]\ ,
\ee
where $\Omega_{\rm m,0}$ and $\Omega_{\Lambda,0}$ are the present day
matter and cosmological constant density parameters,
respectively. Since we are assuming a flat model here, we also have
$\Omega_{\rm m,0}+\Omega_{\Lambda,0}=1$.


\subsection{The galaxy overdensity field on the past lightcone}

Following \cite{Yamamotoetal1999}, let us define the galaxy density
field at time $\eta$, with radial and angular positions
$(\chi,\bgam)$, as:
\be
n(\eta,\chi,\bgam) = n_0(\eta)\left[1+\Delta(\eta,\chi,\bgam)\right] \ ,
\label{eq:n1}
\ee
where $n_0(\eta)$ is the mean number density of galaxies at time
$\eta$ and $\Delta(\eta,\chi,\bgam)$ is the density contrast of
galaxies at the position.  Thus on the past lightcone the number
density is written:
\be
n^{\rm LC}(\br) =
\left.n_0(\eta)\left[1+\Delta(\eta,\chi,\bgam)\right]\frac{}{}\right|_{\eta\rightarrow\eta_0-r/c,\chi\rightarrow r}
\ , \label{eq:n2}
\ee
where $\br\rightarrow(r,\bgam)$ and where $\eta_0$ is the present day
conformal time. We can also define another useful quantity, which
gives the background density on the lightcone:
\be n^{\rm LC}_0(\eta) =
\left.n_0(\eta)\frac{}{}\right|_{\eta\rightarrow\eta_0-r/c} \ .
\label{eq:n3}
\ee
Let us next define the overdensity contrast of galaxies on the past
lightcone as:
\be
\DLC(\br) \equiv \frac{n^{\rm LC}(\br)- n_0^{\rm LC}(r) }{n_0^{\rm LC}(r)}
 = \Delta(\eta',\br)  \ .
\label{eq:n5}
\ee
where we have defined $\eta'\equiv \eta_0-r/c$, which gives the
conformal time at which the density field is recorded on the past
lightcone.  Note that the second equality is important and means that
the density contrast on the lightcone can be rewritten in terms of the
density field at time $\eta'$. In what follows we will make extensive
use of this property.

We now introduce the galaxy survey density field as the quantity:
\be
\FLC(\br) \equiv A\left[ n^{\rm LC}(\br)- \alpha \,n^{\rm LC}_{\rm s}(\br)\right] \label{eq:FLC}\ ,
\ee
where $n^{\rm LC}_{\rm s}(\br)$ is a synthetic galaxy catalogue that
is $1/\alpha$ times denser than the original galaxy catalogue on the
lightcone that replicates all of the selection effects of the true
sample, but which contains no intrinsic spatial correlations. $A$ is a
constant to be determined.


\subsection{Harmonic expansion of the density field}

For an observer viewing their past lightcone the survey selection
function can can be separated into an angular mask and radial
selection function. Furthermore, the evolution of structure and the
redshift space distortions will be noticeable as radial effects. In
what follows it will therefore be useful to decompose the density
field into a set of orthogonal spherical and radial modes. To do this
we will make use of the Spherical-Fourier-Bessel (sFB) expansion of a
scalar field \cite[for a discussion,
  see][]{Peebles1973,BinneyQuinn1991,HeavensTaylor1995} and we provide
an overview of the basic formalism in Appendix~\ref{App:SFB}. The
bottom line is that we can express the density field as:
\begin{align}
\Delta(\br,\eta)
& =  \sum_{\ell,m}
Y_{\ell m}(\hat{\br})\Delta_{\,\ell m}(r,\eta)
\ ; 
& \Delta_{\,\ell m}(r) 
& \equiv  (-i)^\ell \int \frac{\dk}{(2\pi)^3} j_{\ell}(kr)
Y^*_{\ell m}(\hat{\bk})\Delta(\eta,\bk)\ ,\label{eq:Dlmr} 
\end{align}
where the $Y_{\ell m}(\hat{\br})$ are (Laplace) spherical harmonics,
$j_{\ell}(x)$ are the spherical Bessel functions and
$\Delta(\eta,\bk)$ is the Fourier transform of the density field at
time $\eta$.


\section{Spatial statistics on the lightcone} \label{sec:LC1:LCtheory}

Now that we know how to write our galaxy density field on the
lightcone and decompose it into angular and radial modes let us turn
to the question of how to compute the spatial correlations on the past
lightcone.


\subsection{Two-point correlation function on the past lightcone}

Let us now compute the expectation of the product of the lightcone
density field at two separate locations. Using the sFB expansion from
\Eqn{eq:Dlmr} and making use of \Eqn{eq:n5}, we find that this can be
written as \cite{Yamamotoetal1999}:
\begin{align}
  \left<\DLC(\br_1)\DLC(\br_2)\right>
  & = \left<\Delta(\eta_1',\br_1)\Delta(\eta_2',\br_2)\right> \nn \\
  & = \sum_{\ell_1,m_1} \sum_{\ell_2,m_2}  Y_{\ell_1 m_1}(\hat{\br}_1)Y^*_{\ell_2 m_2}(\hat{\br}_2)
  \left<\Delta_{\ell_1 m_1}(\eta_1',r_1)\Delta^*_{\ell_2 m_2}(\eta_2',r_2)
\right> \ ,
\end{align}
where the angled brackets denote an ensemble average process.  On
inserting our expression for the harmonic amplitudes, given by
\Eqn{eq:Dlmr}, the above expression can be written as:
\begin{align}
\left<\DLC(\br_1)\DLC(\br_2)\right>  & = 
\sum_{\ell_1,m_1} \sum_{\ell_2,m_2} (-i)^{\ell_1} i^{\ell_2}
Y_{\ell_1 m_1}(\hat{\br}_1)Y^*_{\ell_2 m_2}(\hat{\br}_2)
\int \frac{\diff k_1 k_1^2}{2\pi^2}\int \frac{\diff k_2 k_2^2}{2\pi^2}
j_{\ell_1}(k_1r_1) j_{\ell_2}(k_2r_2) \nn \\
& \hspace{0.5cm} \times \left<
\Delta_{\ell_1 m_1}(\eta_1',k_1)\Delta_{\ell_2 m_2}^*(\eta_2',k_2)
\right> \ . \label{eq:xi2}
\end{align}
Consider now the expectation factor in the above expression and let us
rewrite this in the following way\footnote{By definition we have
$\Delta_{\ell m}(\eta,k) \equiv \int \diff\hat{\bk}\ Y_{\ell
  m}^*(\hat{\bk})\Delta(\eta,\bk) \ .$}:
\begin{align}
  \left<\Delta_{\ell_1 m_1}(\eta_1,k_1)\Delta_{\ell_2 m_2}^*(\eta_2,k_2)\right>  & =
\int \diff\hat{\bk}_1 \diff\hat{\bk}_2 \ Y_{\ell_1 m_1}^*(\hat{\bk_1}) Y_{\ell_2 m_2}(\hat{\bk_2})
\left<
\Delta(\eta_1',\bk_1)\Delta^*(\eta_2',\bk_2) 
\right> \ .
\end{align}
The correlator $\left<\Delta(\eta_1',\bk_1)\Delta^*(\eta_2',\bk_2)
\right>$ obeys the relation:
\be \left<
\Delta(\eta_1',\bk_1)\Delta^*(\eta_2',\bk_2) 
\right> = (2\pi)^3 \DIII(\bk_1-\bk_2)
P(\bk_1,\eta_1',\eta_2')\ , \label{eq:uneqtc}\ee
where $P(\bk_1,\eta_1',\eta_2')$ is the unequal-time correlator
(hereafter UETC). This comes about due to the fact that the field
$\Delta(\eta,\chi,\bgam)$ obeys statistical homogeneity and isotropy,
and hence so does the cross-correlation between different
epochs. Furthermore, the Dirac delta function can be written in the
spherical polar coordinates as:
\be 
\DIII(\bk_1-\bk_2) = \frac{\DI(k_1-k_2)}{k^2_1}\DII(\hat{\bk}_1-\hat{\bk}_2) \ .
\ee
On putting all of this together and on integrating over the Dirac
delta functions we find \Eqn{eq:xi2} becomes:
\begin{align}
\left<\DLC(\br_1)\DLC(\br_2)\right>  & = 
\frac{(2\pi)^3}{4\pi^4} 
\sum_{\ell_1,m_1} \sum_{\ell_2,m_2} (-i)^{\ell_1} i^{\ell_2}
Y_{\ell_1 m_1}(\hat{\br}_1) Y^*_{\ell_2 m_2}(\hat{\br}_2) \int \diff k k^2 j_{\ell_1}(kr_1) j_{\ell_2}(kr_2)
\nn \\
&  \hspace{0.5cm} \times P(k,\eta_1',\eta_2') \ 
\int \diff\hat{\bk} \ Y_{\ell_1 m_1}^*(\hat{\bk})
Y_{\ell_2 m_2}(\hat{\bk}) \ \label{eq:xi3}\ .
\end{align}
The last integral on the right-hand-side can be computed owing to the
orthogonality of spherical harmonics to give:
\be \int \diff\hat{\bk}_1 \ Y_{\ell_1 m_1}^*(\hat{\bk_1})
Y_{\ell_2 m_2}(\hat{\bk_1}) = \deltaK_{\ell_1,\ell_2}\deltaK_{m_1,m_2}
\ee
where $\deltaK$ is the Kronecker delta symbol. On utilising this fact,
\Eqn{eq:xi3} becomes \cite{Yamamotoetal1999}:
\begin{align}
\left<\DLC(\br_1)\DLC(\br_2)\right>  & =  \frac{2}{\pi}
\sum_{\ell,m}  Y_{\ell m}(\hat{\br}_1)Y^*_{\ell m}(\hat{\br}_2)
 \int \diff k k^2 j_{\ell}(kr_1) j_{\ell}(kr_2)  P(k,\eta_1',\eta_2')\ .
\label{eq:xi4}
\end{align}
However, we now recognise that the integral factor has no dependence
on the azimuthal harmonic $m$. On recalling the addition theorem of
spherical harmonics,
\be {\mathcal L}_{\ell}(\mu_{r_1 r_2}) = \frac{4\pi}{2\ell+1}\sum_{m=-\ell}^\ell
Y_{\ell m}(\hat{\br}_1)Y^*_{\ell m}(\hat{\br}_2) \ ,\ee
where $\mu_{r_1 r_2}=\cos\theta_{r_1 r_2}$ and $\theta_{r_1 r_2}$ is
the angle between the two vectors $\hat{\br}_1$ and $\hat{\br}_2$, we
see that \Eqn{eq:xi4} can be simplified to \cite{CastorinaWhite2018}:
\be
\left<\DLC(\br_1)\DLC(\br_2)\right>  = 
\sum_{\ell}  (2\ell+1) {\mathcal L}_{\ell}(\mu_{r_1 r_2})
\xi_{\ell}(r_1,r_2) \ , \label{eq:corr}
\ee
where ${\mathcal L}_{\ell}(x)$ is the Legendre polynomial of order
$\ell$ and where we defined the correlation function multipole
amplitude as:
\be \xi_{\ell}(r_1,r_2) \equiv \int \frac{\dk}{(2\pi)^3}
j_{\ell}(kr_1) j_{\ell}(kr_2) P(k,\eta_1',\eta_2')
\ .  \label{eq:xil}\ee
Note, while \Eqn{eq:corr} is not a new result, see for example
\cite{Yamamotoetal1999} and \cite{CastorinaWhite2018} for similar
expressions, we have, unlike in these earlier works, not yet made any
approximations for the UETC and so this is true in general (see also
\cite{KitchingHeavens2017}). Furthermore, if we average \Eqn{eq:corr}
over all positions with a fixed separation between $\br_1$ and
$\br_2$, and multiply by the Legendre multipole whose argument is
$\hat{\br}_1\cdot\hat{\br}_2$, one arrives at the lightcone
correlation function multipoles (see \cite{Yamamotoetal2000} for a
discussion), but we shall not discuss these further here.


\subsection{The power spectrum on the past lightcone} \label{ss:PkLC}


With this necessary theory in hand, we now turn to the main aim of the
paper, determining the power spectrum of the observed lightcone
density field $F^{\rm LC}(\br)$. Owing to the fact that the derivation
is somewhat involved, we reserve full details to
Appendix~\ref{app:PkEst}. The main result is that for a flux-limited,
galaxy survey, where galaxies are sampled as discrete points, with a
luminosity dependent biasing, we find the observed power spectrum to
be:
\begin{align} 
\left<\FLC(\bq_1)\left[F^{\rm LC}(\bq_2)\right]^*\right> 
&= 
\prod_{a=1}^{2}
\left\{ A \int \dr_a \Theta(\hat{\br}_a) n^{\rm LC}_{\rm Eff}(\br_a)b^{\rm LC}_{\rm Eff}(\br_a)\right\}
\ex^{i\bq_1\cdot\br_1} \ex^{-i\bq_2\cdot\br_2}\left<\Delta^{\rm LC}(\br_1)\Delta^{\rm LC}(\br_2)\right> \nn \\
& 
+ A^2 \int \dr \ex^{i(\bq_1-\bq_2)\cdot\br} \int dL w^2(\br,L) \Theta(\br|L)\phi(L|\br) \ ,
\end{align}
where $n^{\rm LC}_{\rm Eff}(\br_1)$ and $b^{\rm LC}_{\rm Eff}(\br_1)$
are the effective number density and large-scale bias defined in
equations \Eqns{eq:effbias}{eq:effden},
$\Theta(\br|L)=\Theta(\hat{\br})\Theta(\chi|L)$ is the selection
function defined in \Eqn{eq:select}, $w(\br,L)$ is a weighting
function, and $\phi(L|\br)$ is the evolving galaxy luminosity
function. The second term on the right-hand-side is a shot-noise
term.

If we now focus on the first term on the right-hand-side of the above
expression and make use of our expression for the matter correlation
function on the past lightcone, given by \Eqn{eq:xi4}, we can obtain
the following relation:
\begin{align}
\left<F^{\rm LC}(\bq_1)\left[F^{\rm LC}(\bq_2)\right]^*\right>
& =  \prod_{a=1}^{2}
\left\{ A \int \dr_a \Theta(\hat{\br}_a) n^{\rm LC}_{\rm Eff}(\br_a)b^{\rm LC}_{\rm Eff}(\br_a)\right\}
e^{i\bq_1\cdot\br_1}e^{-i\bq_2\cdot\br_2} \nn \\
& \times
4\pi  \sum_{\ell m}  Y_{\ell m}(\hat{\br}_1)Y^*_{\ell m}(\hat{\br}_2)\xi_{\ell}(r_1,r_2)\ . \label{eq:PF5}
\end{align}
We now make further use of the spherical harmonic expansion of the
plane wave from \Eqn{eq:eikr} and on using these in the above
expression we find:
\begin{align}
\left<F^{\rm LC}(\bq_1)\left[F^{\rm LC}(\bq_2)\right]^*\right>
& =  (4\pi)^3 \prod_{a=1}^{2}
\left\{ A \int \dr_a \Theta(\hat{\br}_a) n^{\rm LC}_{\rm Eff}(\br_a)b^{\rm LC}_{\rm Eff}(\br_a)\right\}
\nn \\
& \hspace{-2cm} \times \ 
\sum_{\ell_1 m_1} i^{\ell_1} j_{\ell_1}(q_1r_1) Y_{\ell_1 m_1}(\hat{\bq}_1)Y_{\ell_1 m_1}^*(\hat{\br}_1) \
\sum_{\ell_2 m_2} (-i)^{\ell_2} j_{\ell_2}(q_2r_2) Y^*_{\ell_2 m_2}(\hat{\bq}_2)Y_{\ell_2 m_2}(\hat{\br}_2)
 \nn \\
& \hspace{-2cm} \times \sum_{\ell m}  Y_{\ell m}(\hat{\br}_1)Y^*_{\ell m}(\hat{\br}_2)
\xi_{\ell}(r_1,r_2)\ ,\label{eq:PF6}
\end{align}
where in rewriting the second plane-wave using the harmonic expansion
we have made use of the fact that the complex conjugate can appear on
either of the spherical harmonics.  The advantage of this is that the
volume integrals over $r_1$ and $r_2$ can now be broken up into radial
and surface parts. The only terms that depend on $\hat{\br}_1$ and
$\hat{\br}_2$ are the spherical harmonics and the angular mask terms
$\Theta(\hat{\br})$. For simplicity let us now assume that we are
dealing with the full sky, such that $\Theta(\hat{\br})=1$
everywhere\footnote{In a future publication we will relax this
condition and develop the formalism for incomplete sky coverage.}, and
so on making repeated use of the orthogonality relation, we have:
\begin{align}
\left<F^{\rm LC}(\bq_1)\left[F^{\rm LC}(\bq_2)\right]^*\right> &
=  4\pi
 \prod_{a=1}^{2}
 \left\{ A \int \dr_a  n^{\rm LC}_{\rm Eff}(\br_a)b^{\rm LC}_{\rm Eff}(\br_a)\right\}
\nn \\
&  \hspace{-2.7cm}\times \ 
\sum_{\ell m}  j_{\ell}(q_1r_1) j_{\ell}(q_2r_2)
Y_{\ell m}(\hat{\bq}_1)Y_{\ell m}^*(\hat{\bq}_2)\xi_{\ell}(r_1,r_2)
  \label{eq:PF8}.
\end{align}
Again, on making use of the addition theorem for spherical harmonics
we find that the above expression can be further simplified to:
\be
\left<F^{\rm LC}(\bq_1)\left[F^{\rm LC}(\bq_2)\right]^*\right>
= \sum_{\ell } {\mathcal P}_\ell(q_1,q_2) {\mathcal L}_{\ell}(\cos\theta_{q_1 q_2}) \ee
where we have defined the multipoles of the $F^{\rm LC}$ power spectrum as
\begin{align}
  {\mathcal P}_\ell(q_1,q_2) & \equiv (2\ell+1)
  \prod_{a=1}^{2}
  \left\{ A \int \dr_a  n^{\rm LC}_{\rm Eff}(\br_a)b^{\rm LC}_{\rm Eff}(\br_a)j_{\ell}(q_ar_a)\right\}
      \xi_{\ell}(r_1,r_2)
  \label{eq:PF8b}.
\end{align}
In practice, the above expression for the multipoles is rather
cumbersome. A more useful variation can be obtained by substituting our
expression for $\xi_{\ell}$ from \Eqn{eq:xil} and on rearranging the
order of integration, moving the $k$ integral to the front, we get:
\begin{align}
  {\mathcal P}_\ell(q_1,q_2)
  & =   (2\ell+1)   \int \frac{\dk}{(2\pi)^3} \prod_{a=1}^2 
  \Big\{ A \int \dr_a  n^{\rm LC}_{\rm Eff}(\br_a)b^{\rm LC}_{\rm Eff}(\br_a)
  j_{\ell}(q_ar_a)j_{\ell}(kr_a)\Big\} P(k,\eta_1,\eta_2)\label{eq:PF9} \ .
\end{align}
Some interesting points to note are: first, for the case where the
unequal time correlator is not a separable function of time, the
evaluation of the above expression requires one to compute a
3D-numerical integral. On the other hand, if it is, then the integrals
can be reduced to a set of 2D integrals (we will demonstrate this in
the next section). Second, the observed lightcone power spectrum,
defined $P_{\rm F}(q) \equiv \left<\left|F^{\rm
  LC}(\bq)\right|^2\right>$, can be obtained from the above equations
by setting $\bq_1=\bq_2$, or equivalently $\theta_{q_1 q_2}=0$,
whereupon ${\mathcal L}_{\ell}(1)=1$, and we have a sum over all
$\ell$ of ${\mathcal P}_\ell(q_1,q_1)$. The final expression for the
real-space monopole is thus:
\begin{align}
  P_{\rm F}(q) & = \sum_{\ell }
  (2\ell+1) \int \frac{\dk}{(2\pi)^3}
  \prod_{a=1}^2
  \Big\{ A\int \dr_j  n^{\rm LC}_{\rm Eff}(\br_a)b^{\rm LC}_{\rm Eff}(\br_a)
  j_{\ell}(qr_a)j_{\ell}(kr_a)\Big\} P(k,\eta_1,\eta_2) + P^{\rm LC}_{\rm shot}\ , \label{eq:PfFull}
\end{align}
where the shot-noise contribution to the observed power spectrum is
given by \cite{Percivaletal2004a}:
\be
P^{\rm LC}_{\rm shot} \equiv A^2 \int \dr  \int^{\infty}_{\Lmin(\br)} dL w^2(\br,L) \phi(L|\br) \label{eq:Pshot}\ .
\ee
It is also interesting to note that, $\left<F^{\rm
  LC}(\bq_1)\left[F^{\rm LC}(\bq_2)\right]^*\right>$ does not vanish
for $\theta_{q_1 q_2}\ne0$. This arises due to the fact that for the
lightcone observer, homogeneity is broken and so there is no Dirac
delta function. However, isotropy is not broken.

\Eqn{eq:PfFull} is one of the main results of this paper -- it
includes the evolution of the galaxy luminosity functions, the
luminosity dependence of galaxy bias, an optimal weighting function
and the shot noise. While some aspects are familiar
\cite{Yamamotoetal1999,Percivaletal2004a}, this result has not been
presented elsewhere. Furthermore, if one were to simplify things by
removing the luminosity dependence of clustering, evolution of the
GLF, and neglect the shot noise one would find that our final analytic
expression for the lightcone power spectrum differs in its final form
from that of \cite{Yamamotoetal1999} (see later for more discussion).


\section{Approximate forms}\label{sec:approx}

We now discuss some important approximations.


\subsection{Time separability of the UETC}

In \Eqn{eq:linbias} we assumed that the galaxies are related to the
matter via a linear bias.  Let us now suppose that the UETC for the
matter can be written as a separable function of time or more
generally as a set of functionals of $P(k,\eta_0)$ each of which has
time separable behaviour. Hence, let us consider the ansatz:
\be P(k,\eta_1,\eta_2) = \sum_{a,b}
G_a(\eta_1,\eta_0)G_b(\eta_2,\eta_0){\mathcal F}_{ab}[P(k,\eta_0)](k) \ , \label{eq:Gen}\ee
where the functions $G_a(\eta,\eta_0)$ give the amplification at time
$\eta$ relative to some fiducial time $\eta_0$ and the functionals
${\mathcal F}_{ab}$ only involve integrals over the equal time
correlator. We note that the above form may be justified on the
grounds that the UETC is symmetric in time,
i.e. $P(k,\eta_1,\eta_2)=P(k,\eta_2,\eta_1)$ and by the fact that this
form will capture a wide range of perturbative expansion schemes.  For
example, in linear theory \Eqn{eq:Gen} has the form:
\be P(k,\eta_1,\eta_2) = D(\eta_1)D(\eta_2)P_{\rm Lin}(k,\eta_0) \ ,\ee
where $G_1(\eta_1,\eta_0)=D(\eta_1)$ are growth factors relative to
time $\eta_0$ and ${\mathcal F}_{ab}[P(k,\eta_0)](k)=P_{\rm
  Lin}(k,\eta_0)$ is the linear theory power spectrum at time
$\eta_0$. Another example, is Eulerian perturbation theory, where at
`1-loop' order the nonlinear power spectrum can (from the commonly
used and accurate Einstein–de Sitter approximation) be expressed in
the form of \Eqn{eq:Gen}. In this case
$G_a(\eta,\eta_0)=\left[D(\eta)\right]^a$ and ${\mathcal
  F}_{13}=P_{13}$, ${\mathcal F}_{22}=P_{22}$ and ${\mathcal
  F}_{12}={\mathcal F}_{21}=0$, giving \cite{KitchingHeavens2017}:
\begin{align}
P(k,\eta_1,\eta_2) & = 
D(\eta_1)D(\eta_2) \Plin(k) + D^2(\eta_1)
D^2(\eta_2) P_{22}(k) \nonumber \\
& + \left[ D^3(\eta_1) D(\eta_2) + D(\eta_1) D^3(\eta_2)\right] P_{13}(k) 
\ . \label{eq:uneqtc1loop}
\end{align}

Focusing on the general separable form, if we insert \Eqn{eq:Gen} into
\Eqn{eq:PF9} then, for the case of $q_1=q_2$ we see that:
\begin{align}
  {\mathcal P}_\ell(q,q)
  & =   (2\ell+1) A^2 \sum_{a,b}
  \int \frac{\dk}{(2\pi)^3}  {\mathcal F}_{ab}[P(k,\eta_0)](k)\nn\\
  & \hspace{0cm}\times 
  \!\!\!\prod_{\alpha=\{a,b\}}\!\!\left\{\int \dr  n^{\rm LC}_{\rm Eff}(\br)b^{\rm LC}_{\rm Eff}(\br)
    G_{\alpha}(\eta,\eta_0)j_{\ell}(qr)j_{\ell}(kr) \right\}
  \label{eq:PF10} \ .
\end{align}
Notice that the final product of integrals are not dependent upon one
another, and this makes the evaluation of the above expression a sum
over 2D integrals. With this result in hand, we see that \Eqn{eq:PfFull}
can be written in the compact form:
\be
P_{\rm F}(q)
=   \sum_{a,b}
\int \frac{\dk}{(2\pi)^3}  {\mathcal F}_{ab}[P(k,\eta_0)](k) {\mathcal R}_{ab}(q,k)+
P^{\rm LC}_{\rm shot}\ .\label{eq:PF11}
\ee
where the kernel window function can be written
\be {\mathcal R}_{ab}(q,k)\equiv
\sum_\ell(2\ell+1) A^2 W_{\ell}^{(a)}(q,k)W_{\ell}^{(b)}(q,k) \ , \label{eq:window_func}
\ee
and we also have defined
\be
W^{(a)}_{\ell}(q,k)\equiv  \int \dr  n^{\rm LC}_{\rm Eff}(\br)b^{\rm LC}_{\rm Eff}(\br)
G_a(\eta,\eta_0)j_{\ell}(qr)j_{\ell}(kr) \label{eq:wina}\ .
\ee
Note, as mentioned at the end of \S\ref{ss:PkLC}, if we neglect the
luminosity dependence of the tracers clustering and shot noise, then
our approximate expression differs from the expression presented in
\cite{Yamamotoetal1999}. That result involves three nested integrals
with the innermost argument being a complicated set of cosine
integrals, which formally introduces a fourth embedded integral.  That
is not to say that the two results are discrepant, more that they are
different representations of the same thing. Our result for the case
of separable UETC is a sum over 2D integrals with products of
spherical Bessel functions, whereas their result involves 4D
integrals. We find the former representation more appealing due to the
reduction in complexity of the required integrals and the fact that,
as we will show, the sum over multipoles converges rapidly on
large-scales.

\begin{figure*}
\centerline{ \includegraphics[width=15.5cm]{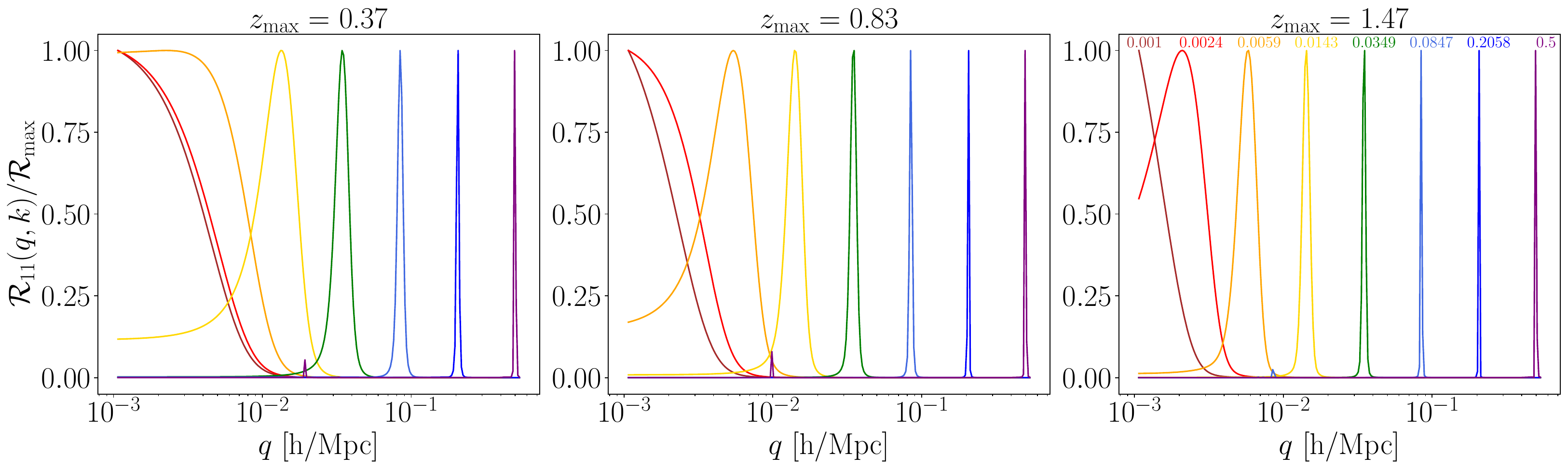} }
\caption{\small{The kernel window function $\mathcal{R}_{11}(q, k)$ of
    \Eqn{eq:window_func} as a function of $q$, for seven values of $k$
    selected to lie in the interval $k\in[0.001, 0.5]\kMpc$, with the
    smallest in dark red and the largest in purple (see annotated
    values in the right-most plot). The panels from left to right show
    the results for volume limited surveys with maximum redshifts of
    $z_{\rm max}=\{0.37,0.83,1.47\}$, respectively, corresponding to a
    maximal radial distance of $\{1, 2, 3\}h^{-1}$Gpc in our chosen
    cosmology.  We see that for scales much smaller than the survey
    volume the window functions take on Dirac delta function like
    behaviour: $\mathcal{R}_{11}(q, k) \approx \delta^D(q - k)$. }
\label{fig:lightcone1:r_integral_plots}}
\end{figure*}


\Figure{fig:lightcone1:r_integral_plots} shows the behaviour of the
kernel window function of \Eqn{eq:window_func}, normalised by its
maximum value. Here we present the case where there is no luminosity
dependent galaxy bias, such that $b(L|r)=1$, and also for the linear
theory UETC.  We have also made the further simplifying assumption of
a volume limited survey with a constant comoving number density of
galaxies $n^{\rm LC}_0(r) = 1$ up to a scale $r_{\rm max}$ and then a
vanishingly small density thereafter. The three panels from left to
right show the results for increasing survey volume. We see that on
small scales (i.e. when $q r_{\rm max} \gg 1$ and $k r_{\rm max} \gg
1$) ${\mathcal R}$ becomes highly spiked and so exhibits delta
function like behaviour where $k\approx q$.  Not too unsurprisingly,
this mimics the window function $G(k-q)$ described in the FKP method
\cite{Feldmanetal1994}.  These plots demonstrate an important point,
which is that when we compute $P_{\rm F}(q)$, to avoid wasting large
amounts of computational time evaluating the $k$ integral in regions
where the integrand is vanishingly small, we will make the bounds $q$
dependent.


\subsection{The large-scale limit: $qr_{\rm max}\ll1$}

Let us consider the asymptotic limit of our full expression for the
power spectrum observable given by \Eqn{eq:PfFull} in the limit where
$qr_{\rm max}\rightarrow 0$. In this case, we notice that
\be \lim_{qr_{\rm max}\rightarrow 0} j_{l}(qr) = \delta^{K}_{l,0}\ . \ee
The consequence of this is that the sum over multipoles in
\Eqn{eq:PfFull} collapses and if we reorder the integrals so that the
$k$-space integral is the innermost one, we see that we have:
\begin{align}
  \lim_{qr_{\rm max}\rightarrow 0} P_{\rm F}(q) & = \prod_{a=1}^2
  \left\{A  \int \dr_a   n^{\rm LC}_{\rm Eff}(\br_a) 
 b^{\rm LC}_{\rm Eff}(\br_a)\right\}
\Sigma(r_1,r_2) + P^{\rm LC}_{\rm shot},
 \label{eq:PfFull2}
\end{align}
where
\be
\Sigma(r_1,r_2) \equiv \int \frac{\dk}{(2\pi)^3} j_{0}(kr_1)j_{0}(kr_2) P(k,\eta_1,\eta_2)\ .
\ee
Consequently, the observed power spectrum on the past lightcone will
asymptote to a constant amplitude on large scales. We have not seen
this result derived elsewhere.


\subsection{The small-scale limit: $qr_{\rm max} \gg 1$}\label{ssec:Pkss}

Conversely, in the small-scale limit, where $q r_{\rm max} \gg 1$ we
have seen from Figure \ref{fig:lightcone1:r_integral_plots} that the
function ${\mathcal R}_{ab}(q,k)$ takes on the behaviour like that of
Dirac's delta function, such that ${\mathcal R}_{ab}(q,k)\propto
\delta^{D}(k-q)$. If the functionals of the matter power spectrum
${\mathcal F}_{ab}[P(k,\eta_0)](k)$ are slowly varying over the scale $q$, then
we can write:
\be 
\lim_{qr_{\rm max}\rightarrow \infty} P_{\rm F}(q)
\approx \sum_{ab} {\mathcal F}_{ab}[P(k,\eta_0)](q) \int \frac{\dk}{(2\pi)^3} \mathcal{R}_{ab}(q, k)
+P^{\rm LC}_{\rm shot}\ . \label{eq:Pfsmall}
\ee
Focusing on the last integral on the right-hand-side of the above
expression we see that from \Eqns{eq:window_func}{eq:wina} this is:
\begin{align}
\int \frac{\dk}{(2\pi)^3} \mathcal{R}_{ab}(q, k)
& = A^2\sum_{\ell} (2 \ell + 1) 
\int \frac{\dk}{(2\pi)^3} \nn \\
& \hspace{-2.0cm}\times 
\prod_{\alpha=\{a,b\}}\left\{\int \dr  n^{\rm LC}_{\rm Eff}(\br)b^{\rm LC}_{\rm Eff}(\br)
G_{\alpha}(\eta,\eta_0)j_{\ell}(qr)j_{\ell}(kr) \right\}\ .
\label{eq:R2}
\end{align} 
By expanding out the product, we see that we can reorder the integrals
so that the $k$-integral is computed first. On recalling the
orthogonality relation of the spherical Bessel functions,
\be
\int \frac{\dk}{(2\pi)^3} j_{\ell}(kr_1) j_{\ell}(kr_2) = \frac{1}{4\pi  r_2^2} \delta^D(r_1 - r_2) \ ,
\ee
we see that the integral over $r_2$ can be done trivially to yield:
\begin{align}
\int \frac{\dk}{(2\pi)^3} \mathcal{R}_{ab}(q, k)
& = A^2 \int \dr  \left[n^{\rm LC}_{\rm Eff}(\br)b^{\rm LC}_{\rm Eff}(\br)\right]^2
 G_{a}(\eta,\eta_0)G_{b}(\eta,\eta_0)\sum_{\ell} (2 \ell + 1)j_{\ell}^2(qr) \nn \\
 & = A^2 \int \dr  \left[n^{\rm LC}_{\rm Eff}(\br)b^{\rm LC}_{\rm Eff}(\br)\right]^2
 G_{a}(\eta,\eta_0)G_{b}(\eta,\eta_0)
  \label{eq:R3} \ ,
\end{align} 
where the final equality in the above expression follows from the fact
that an infinite sum over the squares of spherical Bessel functions is
unity \cite[result 10.1.50]{AbramowitzStegun1970}:
\be
\sum_{\ell}^{\infty} (2 \ell + 1) j^2_{\ell}(x) = 1 \ .
\ee
Thus we see that \Eqn{eq:Pfsmall} tends to the form:
\begin{align}
  P_{\rm F}(q) & \approx A^2 \sum_{ab} {\mathcal F}_{ab}[P(k,\eta_0)](q)
  \int \dr \left[n^{\rm LC}_{\rm Eff}(\br)b^{\rm LC}_{\rm Eff}(\br)\right]^2 G_a(\eta)G_b(\eta)
  +P^{\rm LC}_{\rm shot} \nn \\
  & \approx A^2 \int \dr \left[n^{\rm LC}_{\rm Eff}(\br)b^{\rm LC}_{\rm Eff}(\br)\right]^2
  P(q,\eta) +P^{\rm LC}_{\rm shot}\ . \label{eq:PF12}
\end{align}
By making an appropriate choice for $A$, for example
\be
\frac{1}{A^2} \equiv  \int \dr \left[n^{\rm LC}_{\rm Eff}(\br)\right]^2 \ ,
\ee
we see that in the limit where the the radial weight functions are
very broad and smooth, such that the $W_{\ell}^{(a)}(q,k)$ are very
narrowly peaked, the lightcone power spectrum is approximately given
by the equal time correlator weighted by the square of the selection
function averaged over the radial extent of the survey.  We note that
this result for the small-scale limit agrees with that found by
\cite{Yamamotoetal1999} (see their Eq (16)). However, we note that how
we take the limit differs in detail, owing to the differences in our
final expressions $P_{\rm F}(q)$.


\begin{figure}
  \centerline{ \includegraphics[width=9cm]{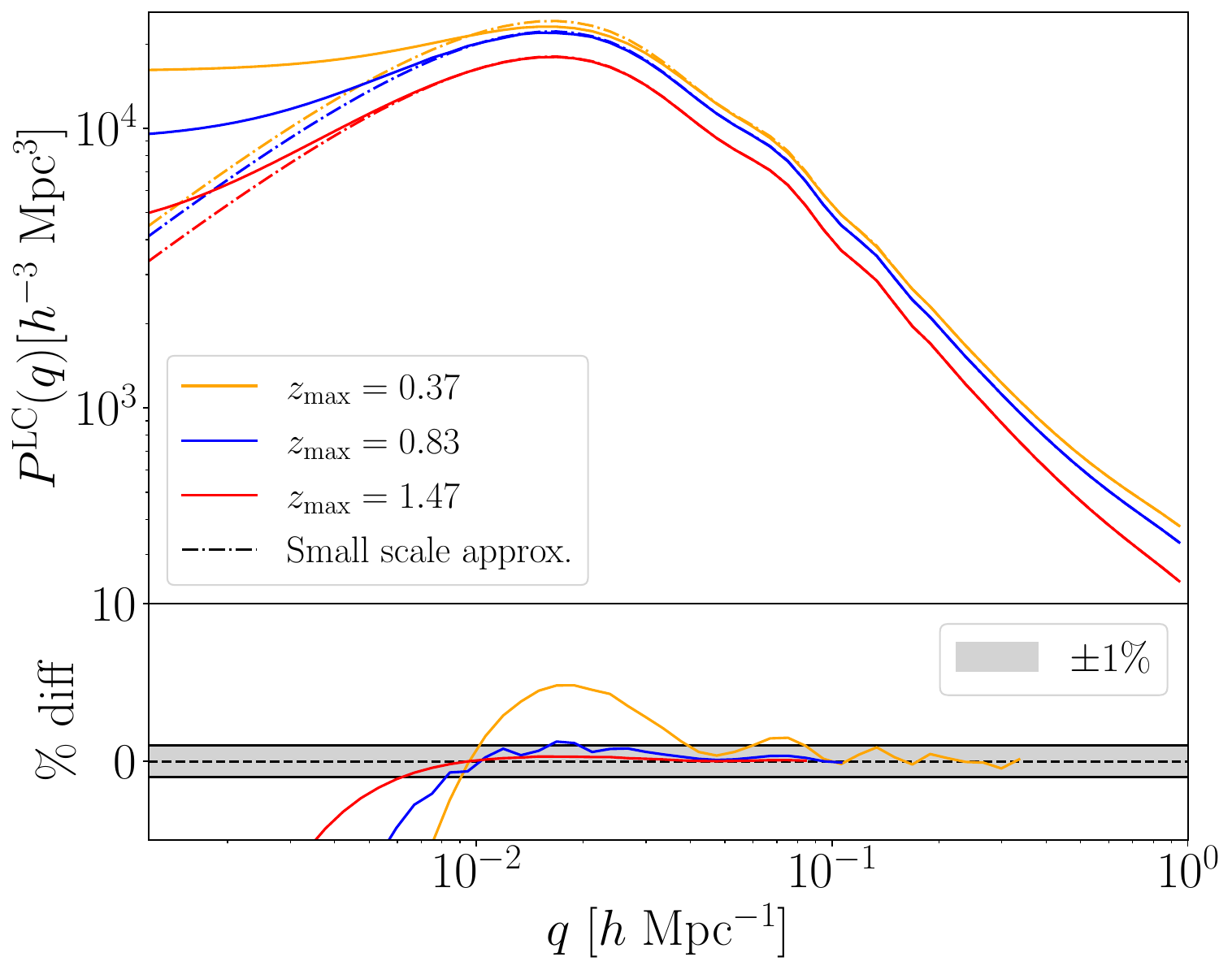}}
  \vspace{-0.2cm}
\caption{\small Upper: A plot of the window-convolved nonlinear power
    spectrum on the lightcone for an all sky survey reaching out to
    increasing redshift depths (corresponding to brighter magnitude
    cuts). The solid lines show the unequal time model
    (\Eqn{eq:PF11}), while the dot-dashed lines show the small
    scale approximation (\Eqn{eq:PF12}). The lower plot shows the percentage difference
    between calculating the power spectrum using the full formula versus using
    the small scale approximation. We see that
    these two formulas reach convergence to well within $1\%$ on small
    scales, where the approximate formula is used to reduce
    computation time.  } \label{fig:lightcone1:UETC_vs_small_approx}
\end{figure}


In \Figure{fig:lightcone1:UETC_vs_small_approx} we plot the window-convolved
power spectrum on the past lightcone for an all-sky survey, out to increasing redshift
depths, using the full expression of \Eqn{eq:PF11} (solid lines). We also plot 
the small-scale approximation given by \Eqn{eq:PF12} with the dash-dotted lines, with
the lower panel of the plot showing the percentage difference between the two
methods. We see that for the cases considered, for $q>0.1\kMpc$, there is $\lesssim
1\%$ difference between the two formulae, and so in this region we can
safely use the approximated formula, which will drastically save on
computational time. We also see that the approximation performs worse
for shallower survey depths, where one has to calculate the full formula
to a higher value of $k$, relative to a deeper survey, before being able
to safely switch over to this approximation.


\subsection{Evaluating theoretical models at a fixed time}

In many large-scale structure analyses, it is common to find that the
theoretical models are evaluated at the mean redshift of the sample,
calculated via \footnote{For an alternative definition of the
effective time (redshift) see \cite{CastorinaWhite2018}}:
\begin{align}
  z_{\mathrm{mean}} &
  = \frac{\int \diff z (\diff V/\diff z)  \; z \; n^{\rm LC}_{\rm Eff}(z)b^{\rm LC}_{\rm Eff}(z)}
  {\int \diff z (\diff V/\diff z) \; n^{\rm LC}_{\rm Eff}(z)b^{\rm LC}_{\rm Eff}(z)} \ . \label{eq:z_mean}
\end{align}
Alternatively, one can instead use an {\em effective survey-fixed
  redshift}. On inspection of \Eqn{eq:PF11} we see that, for a general
nonlinear model, this can not be realised in detail and that the
presence of nonlinear terms with different time dependence violates
this approximation. To emphasise this point more completely, let us
work with the observed power spectrum in the small-scale limit
approximation of \S\ref{ssec:Pkss} given by \Eqn{eq:PF12}. Let us take
this simplified form and, neglecting shot-noise, assume that there
\emph{is} an effective time at which we have the following equality:
\be
P_{\rm F}(q) \approx \int \dr w(r) P(q,\eta)
\stackrel{?}{\approx} P(q,\eta_{\rm eff}) \ ,
\ee
where $w(r)=A^2\left[n^{\rm LC}_{\rm Eff}(\br)b^{\rm LC}_{\rm
    Eff}(\br)\right]^2$ and a super-script question notates that we
are asking whether the conjecture is true. On making repeated use of
\Eqn{eq:Gen} we find:
\begin{align}
\sum_{a,b}\int \dr w(r) G_a(\eta,\eta_0)G_b(\eta,\eta_0){\mathcal F}_{ab}[P(k,\eta_0)](k) \stackrel{?}{\approx} \sum_{a,b}
G_a(\eta_{\rm eff},\eta_0)G_b(\eta_{\rm eff},\eta_0){\mathcal F}_{ab}[P(k,\eta_0)](k)
\label{eq:etaeff}
\end{align}
For the case of linear theory, where $a=1$, $b=1$,
$G_1(\eta,\eta_0)=D(\eta,\eta_0)$ and ${\mathcal F}_{ab}[P(k,\eta_0)](k)=P_{\rm
  Lin}(k,\eta_0)$, we see that a `Linear' effective time $\eta_{\rm
  eff}^{\rm Lin}$ can be found that will satisfy \Eqn{eq:etaeff} if we
can numerically solve the relation:
\be D^2(\eta_{\rm eff}^{\rm Lin}) = \frac{\int \dr  \left[n^{\rm LC}_{\rm Eff}(\br)b^{\rm LC}_{\rm
    Eff}(\br)\right]^2 D^2(\eta,\eta_0)}
    {\int \dr \left[n^{\rm LC}_{\rm Eff}(\br)b^{\rm LC}_{\rm
    Eff}(\br)\right]^2} \ . \label{eq:etaeff2}\ee
We note that this result is related, but not identical, to the
effective redshift noted in \cite{CastorinaWhite2018}. However, what
is now also clear, is that as soon as we add in any additional
nonlinear terms, they would obey a different set of equations.  Hence,
we conclude that there is {\em no effective time which would satisfy
  both the linear and nonlinear evolutionary terms} allowing for a
single effective time.

We can get an idea of the error that is incurred by making the unique
time approximation by considering the case of standard perturbation
theory at the 1-loop level. In this case, if we set $\eta_{\rm eff}$
in accordance with \Eqn{eq:etaeff2}, then the first term in the
left-hand and right-hand-side expansions of \Eqn{eq:etaeff} are
guaranteed to vanish. The next terms which come from the loop
corrections would thus incur the error:
\begin{align}
   P_{\rm Err}(k,\eta_{\rm eff}^{\rm Lin})  & =
   \left[P_{13}(k,\eta_{\rm eff}^{\rm Lin})+P_{22}(k,\eta_{\rm eff}^{\rm Lin})\right] \nn \\
   & \times \left\{1- 
   \frac{\int \dr  \left[n^{\rm LC}_{\rm Eff}(\br)b^{\rm LC}_{\rm
    Eff}(\br)\right]^2 [D(\eta)/D(\eta_{\rm eff}^{\rm Lin})]^4}
    {\int \dr \left[n^{\rm LC}_{\rm Eff}(\br)b^{\rm LC}_{\rm
    Eff}(\br)\right]^2}\right\} \ .
\end{align}
If one could guarantee that, on a given scale, a nonlinear correction
of a certain order was dominant, then one could find a new effective
time to evaluate the theory at. For example, if there is a scale where
the 1-loop contribution is dominant and the linear and 2-loop
corrections are negligible, then one could evaluate the theory at the
1-loop effective time, $\eta_{\rm eff}^{\rm 1-loop}$, defined through
solving the relation:
\be D^4(\eta_{\rm eff}^{\rm 1-loop}) = \frac{\int \dr  \left[n^{\rm LC}_{\rm Eff}(\br)b^{\rm LC}_{\rm
    Eff}(\br)\right]^2 D^4(\eta,\eta_0)}
    {\int \dr \left[n^{\rm LC}_{\rm Eff}(\br)b^{\rm LC}_{\rm
    Eff}(\br)\right]^2} \ . \label{eq:etaeff4}
\ee
Iterating on this logic, one would thus need a set of effective times
for all the scales where the various nonlinear components dominate the
clustering signal.

To compare these fixed-time approximations to the full unequal-time correlator model,
we have assumed the simple case of no weights such that $w(L, r) =1$, 
but have modelled the linear bias as evolving, using the Generalised 
Time Dependent (GTD) model of \cite{Clerkin2015}:
\ba
b^{\rm LC}_{\rm Eff}(\br) = c + \frac{b_0 - c}{D^{\mu}(r)} \ .
\ea
By making the parameter choices of $c = 0.58$, $b_0 = 0.83$ and $\mu = 2.23$, 
this function very accurately represents the widely used model of bias found
in \cite{Tinkeretal2010}, which is calibrated using a wide range of $N$-body simulations.
For modelling the number density on the light cone, $n^{\rm LC}_{\rm Eff}(\br)$, 
we use an evolving Schechter function which we discuss further in \S\ref{ss:nrevo}.


\begin{figure*}
  \centerline{ \includegraphics[width=15cm]{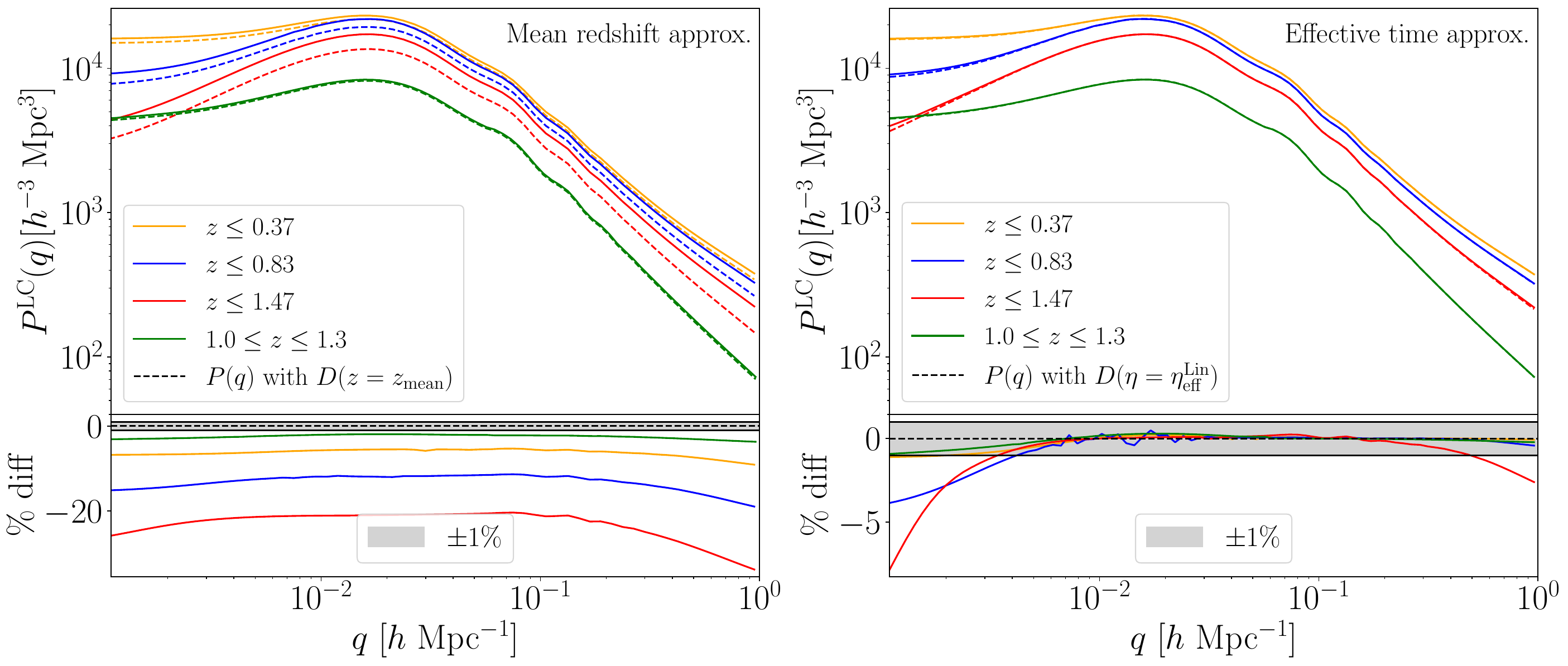}}
  \vspace{-0.2cm}
\caption{\small Theoretical predictions of the mater power spectrum on
  the lightcone, as a function of wavenumber for full-sky surveys of
  different depths (see plot legend). {\bf Left upper panel:}
  Comparison of the full calculation of \Eqn{eq:PF11} (with the
  nonlinear UETC model) with the mean redshift approximation of
  \Eqn{eq:z_mean}, where the power spectrum is evaluated with the
  growth factor fixed at $D(z = z_{\rm mean})$. The solid lines
  represent the full calculation and the dashed lines the
  approximation. {\bf Right upper panel:} same as left panel, except
  the comparison is with the effective time approximation $D(\eta =
  \eta_{\rm eff}^{\rm Lin})$ from \Eqn{eq:etaeff2}.  The lower section
  of each panel shows the percentage difference between the full
  calculation and the approximation. The shaded region shows $\pm 1\%$
  deviation. } \label{fig:lightcone1:UETC_vs_rmean_reff_bias}\vspace{-0.4cm}
\end{figure*}


In \Figure{fig:lightcone1:UETC_vs_rmean_reff_bias} we illustrate the
impact of using the different approaches to a fixed-time
approximation. In both of the plots, we show the unequal-time power
spectrum model (solid lines) for increasing survey depth.  This is
calculated using \Eqn{eq:PF11} on large scales, and uses the
small-scale approximation of \Eqn{eq:PF12} for scales well within the
scale of the survey window function.  In the left hand plot, we
compare this with the mean redshift approximation of \Eqn{eq:z_mean},
and in the right hand plot we instead use the effective redshift (or
time) approximation calculated from \Eqn{eq:etaeff2}, with both
approximations represented by dashed lines\footnote{Note that for each
of these approximations, the only quantity we are changing when
compared to the full UETC model is to evaluate the growth factor at
some fixed redshift.}. The lower plots in each case show the
percentage difference between the full model and the given
approximation, with the shaded region showing the $\pm 1\%$
threshold. We see that using the mean redshift approximation results
in a $k$-dependent bias of the power spectrum amplitude, which
increases with severity as the survey depth increases. At intermediate
scales where there is minimal convolution with the survey window
function and non-linear effects are small, this offset is roughly
constant due to the fact that the linear power spectrum scales as
$D^2(r)$. We notice that the difference grows larger at smaller scales
where we pick up non-linear corrections, depending on higher powers of
the growth factor. Conversely, from the right hand plot, we see that
using the effective time approximation works very well at all but the
largest and smallest scales, only deviating from sub-percent level
accuracy at these extremities for the deeper survey examples.
Therefore, where a fixed time approximation is required to evaluate
theory, we strongly advocate the use of \Eqn{eq:etaeff2} to do this.


\subsection{Extending the theory to the nonlinear regime} \label{ss:nonlin}

Building on the analysis of the previous sections, it is now
straightforward to extend our model for the power spectrum in the
small-scale limit to the non-linear regime. We do this by
perturbatively expanding the density field in terms of powers of the
square of the growth factor. Using the approximate form of
\Eqn{eq:PF12}, we now substitute in the equal-time version of the UETC
expansion of the density field from \Eqn{eq:uneqtc1loop}, and up to
`2-loops' this, gives us:
\begin{align}
  P_{\mathrm{F}}(q)
  & \approx P_{\mathrm{Lin}}(q) \mathcal{W}^{(2)} + P_{\mathrm{1-loop}}(q)
  \mathcal{W}^{(4)}  + P_{\mathrm{2-loop}}(q) \mathcal{W}^{(6)} , \label{eq:approxpk2loop}
\end{align} 
where for brevity we have made the definition:
\be \mathcal{W}^{(i)} \equiv A^2 \int \dr \left[n^{\rm LC}_{\rm Eff}(\br)b^{\rm LC}_{\rm Eff}(\br)
    \right]^2 D^i(r)\ .  \ee

In summary, our final model for the power spectrum on the past
lightcone is a piecewise combination of two different models: on large
scales, we use the full expression of \Eqn{eq:PF11} with the 1-loop
UETC model of \Eqn{eq:uneqtc1loop}, and on small scales, we use the
approximation of \Eqn{eq:approxpk2loop}. Furthermore, we use
\texttt{CAMB} to generate the linear theory $P_{\mathrm{Lin}}$, and
then FAST-PT \cite{McEwenetal2016}, with $P_{\mathrm{Lin}}$ as input,
to generate the 1-loop terms. Finally, to calculate a proxy for the
`2-loop' terms we compute the nonlinear power spectrum model of
\cite{Takahashietal2012}, and subtract off $P_{\mathrm{Lin}}$ and the
1-loop contribution from FAST-PT. These are all generated at $z=0$ and
then scaled as appropriate using powers of the growth factor $D^i(r)$
as indicated above.


\section{Validation with $N$-body simulation mock catalogues} \label{sec:LC1:mocks}

We now turn to the issue of validating the derived expressions using
galaxy mock catalogues obtained from $N$-body simulations of the past
lightcone. We first detail the $N$-body data that we use, the process
for creating the dark matter lightcones, and then finally the estimator
methodology that we use to measure the power spectrum.


\subsection{D\"ammerung Simulation}

To generate past lightcones we make use of the large-volume simulation
from the ``D\"ammerung Suite'' of runs \cite[for details
  see][]{SmithAngulo2019}, which allows us to measure the power
spectrum over a wide range of scales. In summary, the cosmological
parameters of this run were in accord with the Planck best-fit
\cite{Planck2014XVI}. The exact cosmological parameter values that
were used are: the dark energy equation of state parameters were
$w_0=-1.0$ and $w_a=0.0$; the dark energy density parameter was
$\Omega_{\rm DE}=0.6914$, which, since the cosmological model was
spatially flat gave a matter density $\Omega_{\rm m}=0.3086$; the
physical densities of cold dark matter and baryons were set to
$\omega_c=0.11889$ and $\omega_b=0.022161$, respectively; the
primordial power spectrum spectral index, amplitude and running were
set to $n_{\rm s}=0.9611$, $A_s=2.14818\times 10^{-9}$ and
$\alpha=0.0$, respectively.  The linear matter power spectrum was
computed using {\tt CAMB} \cite{Lewisetal2000}, down to $z=0$.  This
was rescaled back to the $z=49$ using the scale-independent matter
only linear growth factor and the initial conditions were lain down
using an upgraded version of 2LPT \cite{Crocceetal2006}.

The simulation was run using the {\tt Gadget-3} code developed for the
Millennium-XXL simulation\cite{Springel2005,Anguloetal2012}.  The
large-volume run was performed with $N=2048^3$ dark matter particles,
in a comoving box of size $L=3000\Mpc$, yielding a mass per particle
of \mbox{$m_{\rm p}=2.69\times10^{11}\rm{\Msol}$}. Sixty snapshots
were output between $z=49$ and $z=0$, with a hybrid linear-logarithmic
output spacing that matched the Millennium Run I simulation
\cite{Springeletal2005}. The simulation was run on the SuperMUC
machine at the Leibniz Rechnum Zentrum in Garching and the full
particle data storage was $\sim$20 TB.


\subsection{Construction of the past lightcones for dark matter particles}

Full details of our method for construction of the past lightcone will
described separately in Booth et al. (in prep.), but the basic
methodology follows the work of \cite{Mersonetal2013}, differing
though in the details. The main steps of the algorithm can be
summarised as follows:
\begin{itemize}
\item We wish to solve the lightcone crossing equation for each dark
  matter particle in the $N$-body simulation. Concretely, for the
  $k$th particle we want find the value of $t^{\rm c}$ that satisfies
  the equation
\be
\left|\br_k(\chi,\theta,\phi,t^{\rm c})-\br_{\mathcal O}\right|=\int^{t_0}_{t^{\rm c}} \frac{cdt}{a(t)}\ ,
\label{eq:WL1}
\ee
where $\br_k$ is the worldline of the particle, $\br_{\mathcal O}$ is
the observers location and $t^{\rm c}$ gives the coordinate time when
the particle exits the past lightcone.

\item In order to solve the above equation, we need to reconstruct the
  full world line of each particle. We do this in a piece wise fashion
  by using a Taylor expansion up to cubic order in lookback time to
  interpolate the particle positions and velocities between
  neighbouring snapshots. The parameters of the Taylor expansion are
  fixed using the particle positions and velocities at the snapshots.
  For example, the equation for the Cartesian $x$ component of the
  particle world line is:
  \begin{align}
    x(t)
    & \approx \overline{X}-\frac{\Delta V\Delta T}{8}   +\frac{\left(3\Delta X
      -\overline{V}\Delta T\right)}{2}\left(\frac{t-\overline{T}}{\Delta T}\right)
    + \frac{\Delta V\Delta T}{2}  \left(\frac{t-\overline{T}}{\Delta T}\right)^2 \nn \\
    & 
    -2\left[\Delta X-\overline{V}\Delta T\right]
  \left(\frac{t-\overline{T}}{\Delta T}\right)^3 ;
  \label{eq:WL7a}
  \end{align}
where $\overline{T}\equiv (t_{i}+t_{i+1})/2$ and where $\Delta
t=t-\overline{T}_i$ with $t_i$ and $t_{i+1}$ being the lookback times
to the neighbouring snapshots, with $t_i<t_{i+1}$. Also we have
defined $\overline{X}\equiv (x_i+x_{i+1})/2$, $\Delta X\equiv
x_{i+1}-x_{i}$, $\overline{V}\equiv (v_{i+1}+v_{i})/2$, and $\Delta
V\equiv v_{i+1}-v_{i}$, with $(x_{i},v_i)$ and $(x_{i+1},v_{i+1})$
referring to the particle's position and velocity for the neighbouring
snapshots, respectively. In Booth et al. (in prep.) we show that for
our simulation we can do this to an accuracy of $r\lesssim 200\kpc$,
for all particles.
\end{itemize}

We apply the above algorithm to generate 8 different dark matter
particle light cones, where we have set the different observer
locations to be the vertices of a cubical lattice of side
$L/2=1500\Mpc$. The data footprint of each particle light cone is
roughly $\sim2$TB. While these ligtcones are not fully independent, on
scales smaller than $r\lesssim L/4$ they have no repeated structures
and thus for shallower surveys can be considered independent from one
another. The further advantage of doing this for the deeper lightcones
is that one averages over similar structures, but at different
epochs. Nevertheless, in what follows while we will present the mean
statistic averaged over the 8 quasi-independent light cones, we will
compute the errors on the observables using the Gaussian theory
estimates.


\subsection{Constructing the galaxy past lightcone} \label{ss:nrevo}

We would like to be able to explore how well the theory from the
previous sections works for a galaxy sample that would be comparable
to that for the Bright Galaxy Sample (hereafter BGS-like) from DESI or
the 4MOST Cosmology Redshift Survey.  Owing to the fact that the
large-volume ``D\"ammerung run'' does not have sufficient resolution
to resolve the typical halo masses in these surveys, we adopt the
simplified strategy of assuming that the BGS-like galaxies are a
Poisson sampling of the dark matter particles (i.e. are unbiased).
However, we do introduce realistic radial selection functions to
emulate the effects of the flux-limits that we adopt. Our mock galaxy
recipe follows the following steps:

\begin{enumerate}

\item Adopt a magnitude limit, $m_{\rm lim}$, and set bounding
  redshifts $z_{\rm min}$ and $z_{\rm max}$ for the survey.
\item Next, we assume an (evolving) galaxy luminosity function (GLF)
  for the survey. The number density of particles in the dark matter
  past lightcone sets the maximum density of galaxies in the mock
  catalogue, such that:
  \be
  \bar{n}_{\rm max}(r) \ge \int_{\rm L_{\rm min}(r)}^{\infty} \phi(L,r)\diff L \ ,
  \ee
  where $\phi(L,r)$ is the evolving GLF, and in our case we follow the
  work of \cite{Lovedayetal2012} and use a Schechter function form:
  \be
  \phi(L,r) \diff L = \phi^*(r) \left( \frac{L}{L^*(r)} \right)^{\alpha} 
  \exp\left(-\frac{L}{L^*(r)}\right) \frac{\diff L}{L^*(r)} \ ,
  \ee
  where the parameters $\phi^*(r)$ and $L^*(r)$ evolve as a function
  of radial comoving distance and $\alpha$ is fixed for all redshifts
  (for full details see Appendix~\ref{App:Schech}).  Thus for a given
  $\bar{n}_{\rm max}$ we see that this imposes the minimum luminosity
  that a galaxy could have and still make it into the survey.
\item To each dark matter particle in the lightcone we now sample a
  luminosity from the GLF following the methodology outlined
  in Appendix~\ref{App:Mags}.
\item For the specified flux-limit, and for a given particle with
  redshift $z$, we then use \Eqn{eq:Lmin} to determine the minimum
  luminosity that a galaxy could have at that distance and be
  contained in the survey. This step is most efficiently done through
  creating a cubic-spline of this function. We then include or exclude
  the potential mock galaxy (particle) based on whether its luminosity
  satisfies $L\ge\Lmin(r)$.
  
\item Finally, this process is repeated for all of the flux-limit
  samples of interest.

\end{enumerate}
We take 5 magnitude cuts at $m_{\rm{lim}} = \{18,19, 20, 21, 22\}$
(see Appendix~\S\ref{App:Mags} for details on sampling). We note that
these catalogues are thus not independent realisations from one
another, but rather the faintest catalogue, $m_{\rm{lim}} = 18$,
contains a subset of particles from the $m_{\rm{lim}} = 19$ catalogue,
which in itself is a subset of the $m_{\rm{lim}} = 20$ catalogue, and
so on.


\begin{figure}
\centerline{ \includegraphics[width=9cm]{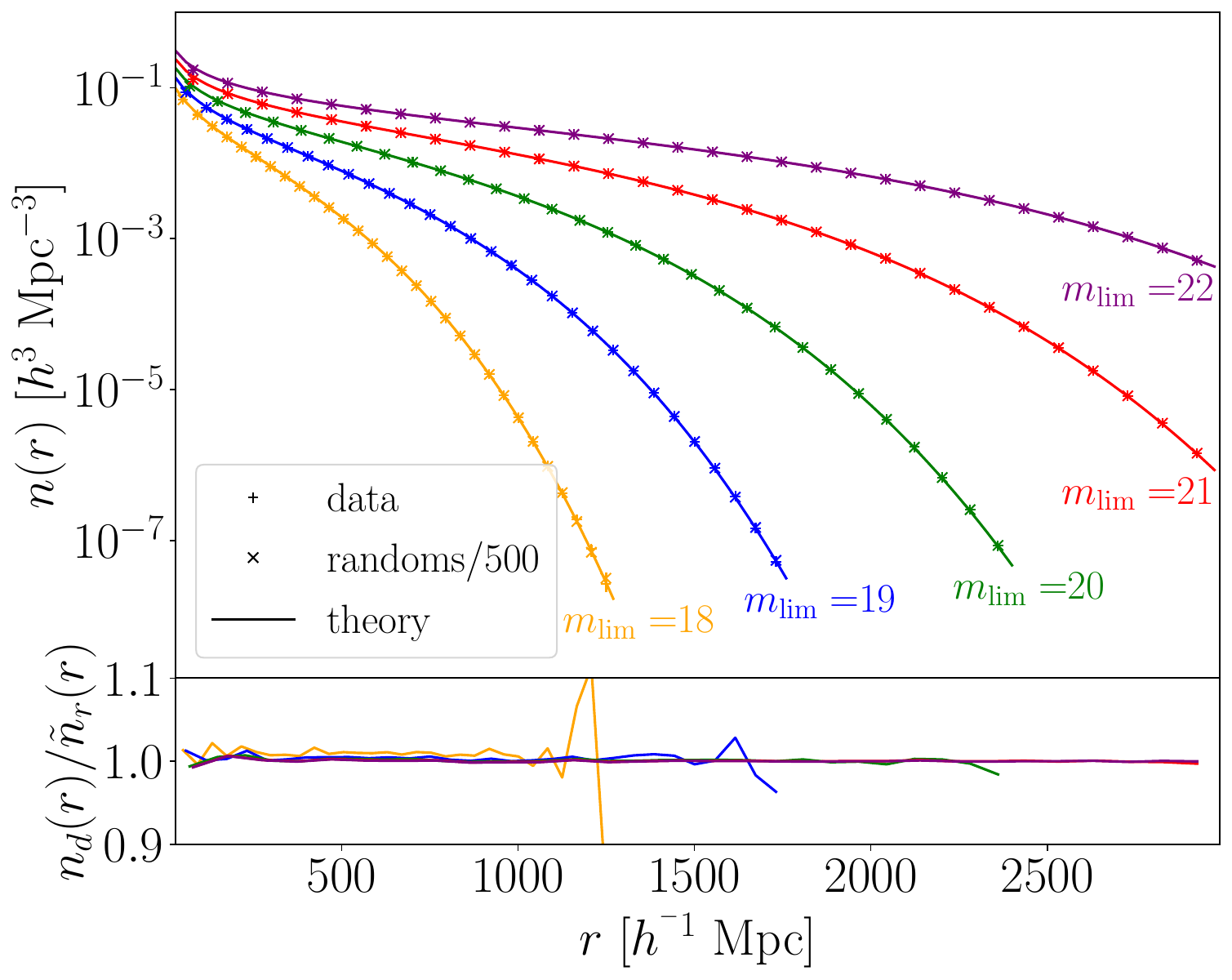}}
\caption{\small Evolution of the mean galaxy density as a function of
  the comoving distance. {\bf Upper panel}: the galaxy number
  densities measured from the mock are shown by the plus symbols, the
  crosses show the same but for the random catalogue scaled by
  $1/R_{\rm multi}$. Different colours indicate the different
  magnitude limits. The solid lines represent the theoretical
  prediction from integrating the evolving Schechter function, as in
  \Eqn{eq:nrLC}. {\bf Lower panel}: the ratio of the number densities
  measured in the mock data with that of the random distributions down
  weighted by $R_{\rm multi}$. } \label{fig:lightcone1:nrdist}
\end{figure}


\begin{figure*}
\centerline{ \includegraphics[width=15cm]{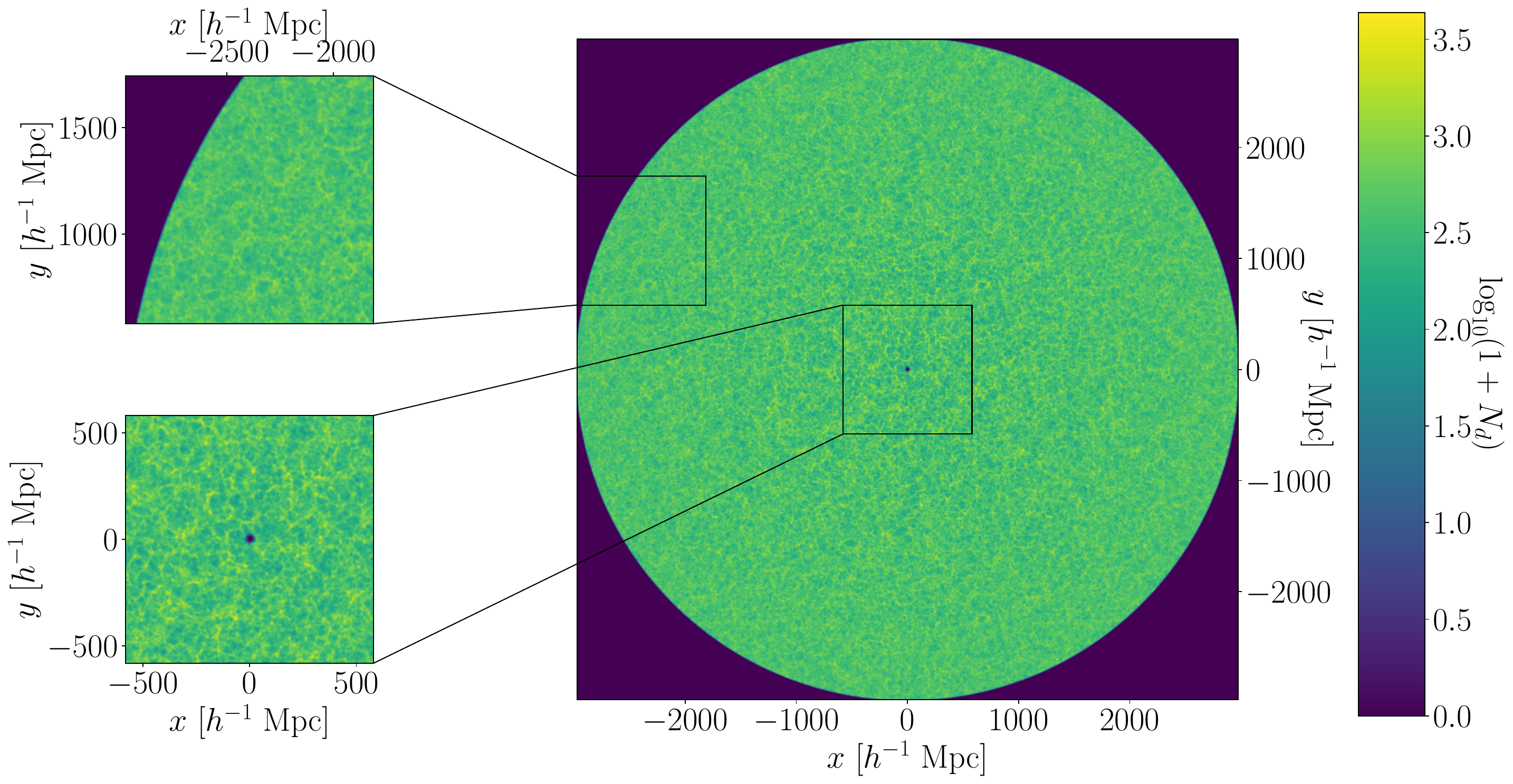} }
\caption{\small{Cross section slice through the centre of the full-sky
    dark matter lightcone, including zoom regions near the edge and
    the centre of the sphere. The cross section is $40\,\Mpc$ thick
    and spans a redshift range of $z=0$ at the centre to $z \approx
    1.45$ at the outer edge. The heatbar at the right shows the
    logarithm to base ten of the particle number per pixel.}
\label{fig:lightcone1:datgrid}} 
\end{figure*}


We also generate unclustered mock random data sets to go alongside
these magnitude-cut catalogues, for use in our power spectrum
measurements. These random catalogues are created by distributing
points uniform-randomly in a sphere of radius $r_{\rm{max}} = 3$ $h^{-1}$Gpc,
and then sampling magnitudes using the same method as above and making
matching apparent magnitude limit cuts -- one small difference though,
is that we omit Step~ii and use the same $L_{\rm min}(r)$ as the
clustered catalogue to sample luminosities with. This process is
iterated for each cut, until we have 5 mock random catalogues, which
match the overall survey geometry of their data counterparts, but with
$N_{\rm{rand}} = R_{\rm multi} N_{\rm{data}}$ for each. In our case,
we set $R_{\rm multi} \simeq 500$, choosing this high value to reduce
the probability of any bias or random error.  Table
\ref{tbl:sim_cats}, presents some of the key details of our full sky
dark matter lightcone and the details of our 5 flux-limited mock
galaxy catalogues, averaged over the 8 realisations.

\Figure{fig:lightcone1:nrdist} shows the resultant number densities of
the different mock galaxy catalogues for both the data and the
randoms. Here we also show the expected theoretical prediction given
by \Eqn{eq:nrLC} as the solid lines and as can be seen the mock and
random catalogues and the theory have exactly the same evolution with
comoving radial distance.

\Figure{fig:lightcone1:datgrid} shows a slice through one of the
full-sky dark matter lightcones and several zoom-in regions. These
panels visually demonstrate how the clustering of the large-scale
structure drops off as on goes from the inner ring to outer rim of the
lightcone sphere.


\begin{table}
\begin{center}
\begin{tabular}{ |c||c|c|c|  }
 \hline
 $m_{\rm lim}$ & Avg. $N_d$ & $z$ range & $\chi$ range [$h^{-1} {\rm Mpc}$] \\
 \hline
 None  & $35.0 \times 10^9$ & $0.01 \leq z \leq 1.45$ & $30 \leq \chi \leq 2974$ \\
 18.0  & $4.6 \times 10^6$ & $0.01 \leq z \leq 0.36$ & $30 \leq \chi \leq 975$ \\
 19.0  & $16.7 \times 10^6$ & $0.01 \leq z \leq 0.53$ & $30 \leq \chi \leq 1390$ \\
 20.0  & $60.0 \times 10^6$ & $0.01 \leq z \leq 0.81$ & $30 \leq \chi \leq 1950$ \\
 21.0  & $212.8 \times 10^6$ & $0.01 \leq z \leq 1.24$ & $30 \leq \chi \leq 2680$ \\
 22.0  & $756.0 \times 10^6$ & $0.01 \leq z \leq 1.45$ & $30 \leq \chi \leq 2974$ \\
 \hline
\end{tabular}
\caption{List of mock galaxy lightcones. Column 1, gives the apparent
  magnitude cut (the case where no limit is taken gives the dark
  matter lightcone); column 2 gives the galaxy/particle count $N_d$
  averaged over the 8 realisations; column 3 gives the `effective'
  redshift range; and column 4 the corresponding comoving
  distance. Note that the upper bound of these ranges signifies where
  $n(r)$ drops 4 orders of magnitude below its peak value, where it is
  $\le2974 h^{-1} {\rm Mpc}$ (see \Fig{fig:lightcone1:nrdist}).}
\label{tbl:sim_cats}
\end{center}
\end{table}


\subsection{The power spectrum estimator}\label{sec:est}


\begin{figure}
\centerline{ \includegraphics[width=9cm]{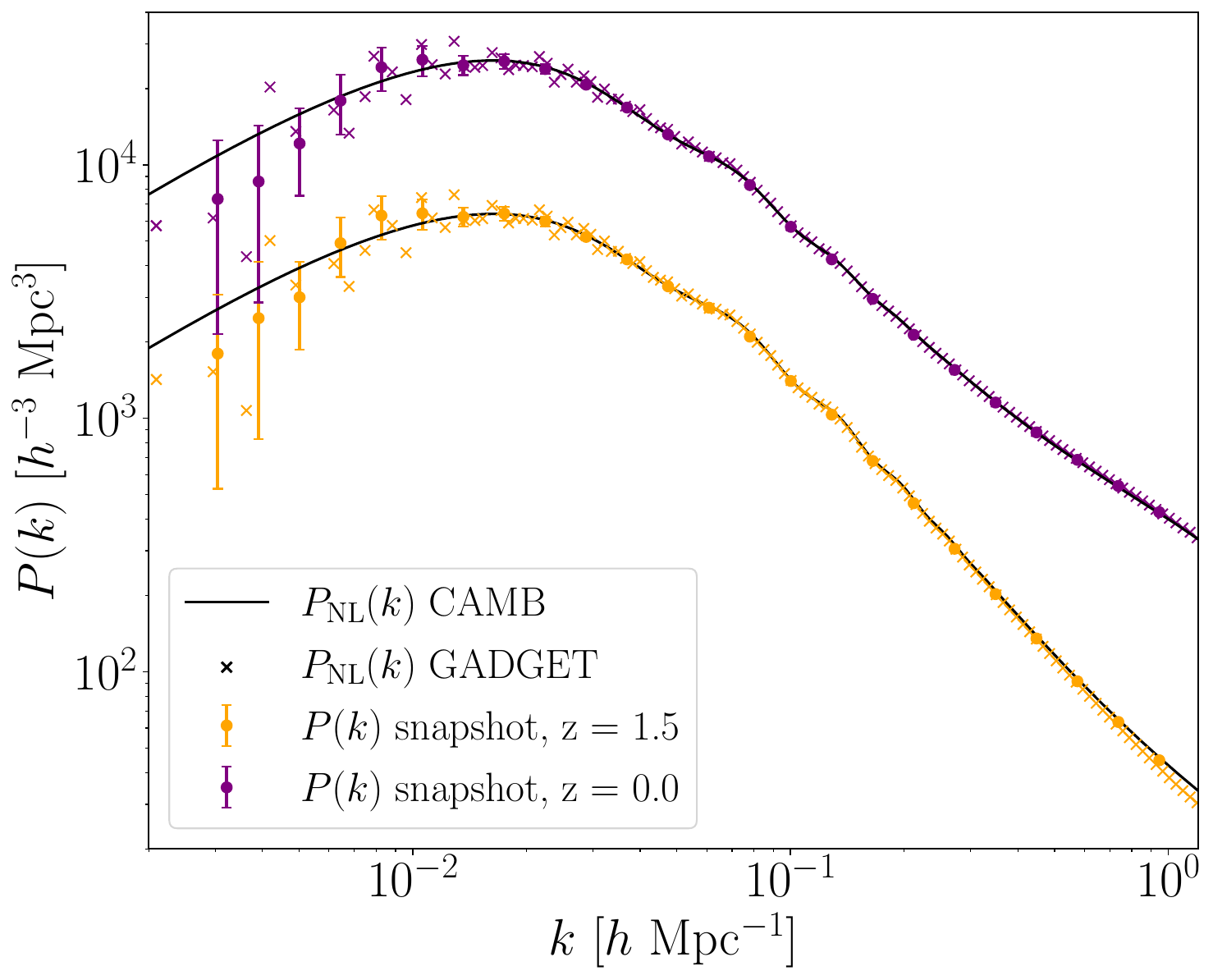} }
\caption{\small{Validation of the matter power spectrum $P(k)$
    estimates from the D\"ammerung simulation snapshots as a function
    of wavenumber. The orange and purple symbols show the results from
    snapshots $z=1.5$ and $z=0$, respectively.  The solid circles show
    the results from our estimator.  The crosses represent the output
    from the \texttt{GADGET-4} code. The black solid lines show the
    nonlinear $P(k)$ predictions from \texttt{CAMB}. }
\label{fig:lightcone1:daemmerung_pk}}
\end{figure}


To estimate the power spectra from each catalogue, we broadly follow
the FFT based method of \cite{Feldmanetal1994} and
\cite{Blakeetal2010}, but with some small variations, which we
summarise below.

\begin{itemize}
\item First, both the mock data and random catalogues were
  interpolated onto a cubical mesh of dimension $N_{\rm grid}=1024^3$,
  using the Triangular-Shaped-Cloud (TSC) mass assignment scheme
  \cite{Hockney1981}. The cubical mesh was selected to be large enough
  to contain the diameter of the dark matter light cone.

\item We then implemented the estimator of \cite{Blakeetal2010} in
  Python 3.6 (Python Software Foundation, https://www.python.org/),
  making use of {\tt NumPy} \cite{Harrisetal2020}, {\tt SciPy}
  \cite{Virtanenetal2019}, {\tt AstroPy} \cite{Astropy2013}, {\tt
    Numba} \cite{Lametal2015}, and the python wrappers for MPI
  \cite{Dalcinetal2011} and FFTW \cite{Gomersall2016}.
  
\item To reduce the effects of small-scale aliasing caused by the
  finite resolution of the grid \cite[see][for a
    discussion]{Jing2005}, we employed the interlacing method of
  \cite{Sefusattietal2016}. In that work, it was demonstrated that,
  through combining the TSC assignment with interlacing, one is able
  to measure the power spectrum in a periodic box to $\lesssim 0.1\%$
  accuracy all the way up to the Nyquist frequency of the grid
  ($k_{\rm Ny}=\pi N_{\rm grid}/L$).

\item For the covariance, we took Eqn~(20) from \cite{Blakeetal2010}:
  \be \left< \delta P(\bk_i) \delta P(\bk_j)\right> =
  \frac{2
    A^4 \sum_{\bk_i, \bk_j} |P(\bk_i)Q(\bk_i - \bk_j) +
    S(\bk_i - \bk_j)|^2 }{N_{\bk_i} N_{\bk_j}} , \label{eq:FKPCov} \ee
  where $N_{\bk_i}$ is the number of modes in a given bin and the
  equations for $Q(\bk)$ and $S(\bk)$ are given by:
  \begin{align}
  Q(\bk) & = \sum W^2(\bx) w^2(\bx) \exp(i\bk \cdot \bx) \ ;\\
  S(\bk) & = \sum W(\bx) w^2(\bx) \exp(i\bk \cdot \bx)\ .
  \end{align}
\end{itemize}
It is worth pointing out that while we have assumed that the
underlying Fourier modes of the galaxy density field are Gaussianly
distributed, \Eqn{eq:FKPCov} shows that the observed power spectrum in
different $k$-bins \emph{does} become correlated, due to the the
survey window function. However, in the limit that the survey window
function is sufficiently compact, centred on each bin, we see that the
diagonal covariance approximation will be accurate on scales that lie
sufficiently within the survey. For larger scales though, this will be
an under-estimate. The former case leads to the following approximate
form for the covariance (FKP Eqn.\,(2.2.6)):
\ba \left< [\delta P(\bk_i)]^2\right> \simeq \frac{2
  A^4}{N_{\bk}}|P(\bk_i)Q(\bk_i) + S(\bk_i)|^2 .
\label{eq:FKPCov2}
\ea
%


To test the accuracy of the code, we first applied it to measure the
power spectrum of the original D\"ammerung Simulation redshift
snapshots, where our results can easily be compared to the output of
\texttt{CAMB} nonlinear power spectra at the specified redshift, as
well as the power measured by \texttt{GADGET-4}
\cite{Springeletal2021}. We perform the measurements in 25
logarithmically spaced bins, where the upper and lower bin edges are
set to match the largest and smallest modes supported by the box,
giving bin edges such that $0.0021\kMpc \leq k \leq 1.07\kMpc$, and
bin width $\log_{10}(\Delta k/[\kMpc]) = 0.108$.

Figure \ref{fig:lightcone1:daemmerung_pk} presents the results from
these tests, where we measure the power spectrum at redshifts $z =
0.0$ and $z = 1.5$, which is the approximate redshift range spanned by
our lightcone simulations. We see excellent agreement between our
measurements and the nonlinear prediction from \texttt{CAMB} and
\texttt{GADGET} across all scales.  At the largest scales, just beyond
the turnover scale, we see a slight increase in power relative to the
linear theory. This we attribute to cosmic variance. As we will in the
next section, this noise feature is also present in the lightcone
measurements. Note that the error bars are based on \Eqn{eq:FKPCov2}
and that we have computed the spectrum all the way to the Nyquist
frequency of the Fourier mesh with no resultant boost in power,
characteristic of aliasing effects. This gives us confidence that the
estimator methodology is working as desired, and we now move on to
tackle the analysis of the lightcone data.

 
\section{Results: power spectrum on the past lightcone} \label{sec:LC1:results}


\subsection{Measurements of the mock lightcone survey catalogues}
 

We now measure the power spectrum for the 8 pseudo-independent
realisations of our lightcone, in the 5 different apparent magnitude
limited mock galaxy catalogues.  This gives us a total of 40 sets of
measurements to perform. As with our testing on the individual
D\"ammerung snapshots, we measure the power in 25 logarithmically
spaced bins, but this time with $k$-range of $0.0042\kMpc < k \leq
0.54\kMpc$. This gives a $k$-bin width of $\log_{10}(\Delta k/[\kMpc])
= 0.0844$. The reduction in the upper wavemode cut-off is due to the
fact that the length of the cubical box being twice the size of the
original D\"ammerung Simulation box. The effect of which is to halve
the Nyquist frequency. In addition, we set the lowest frequency bin to
be twice the fundamental mode of the base simulation $k_{\rm
  f}=2\pi/(1500 h^{-1} \rm{Mpc})\approx0.0042\kMpc$. For wavemodes
below this scale mode discreteness and the box replication is relevant
and we would expect repeated structures to appear on the same
lightcone, despite their `existing' at different epochs.


\begin{figure*}
\centerline{ \includegraphics[width=16cm]{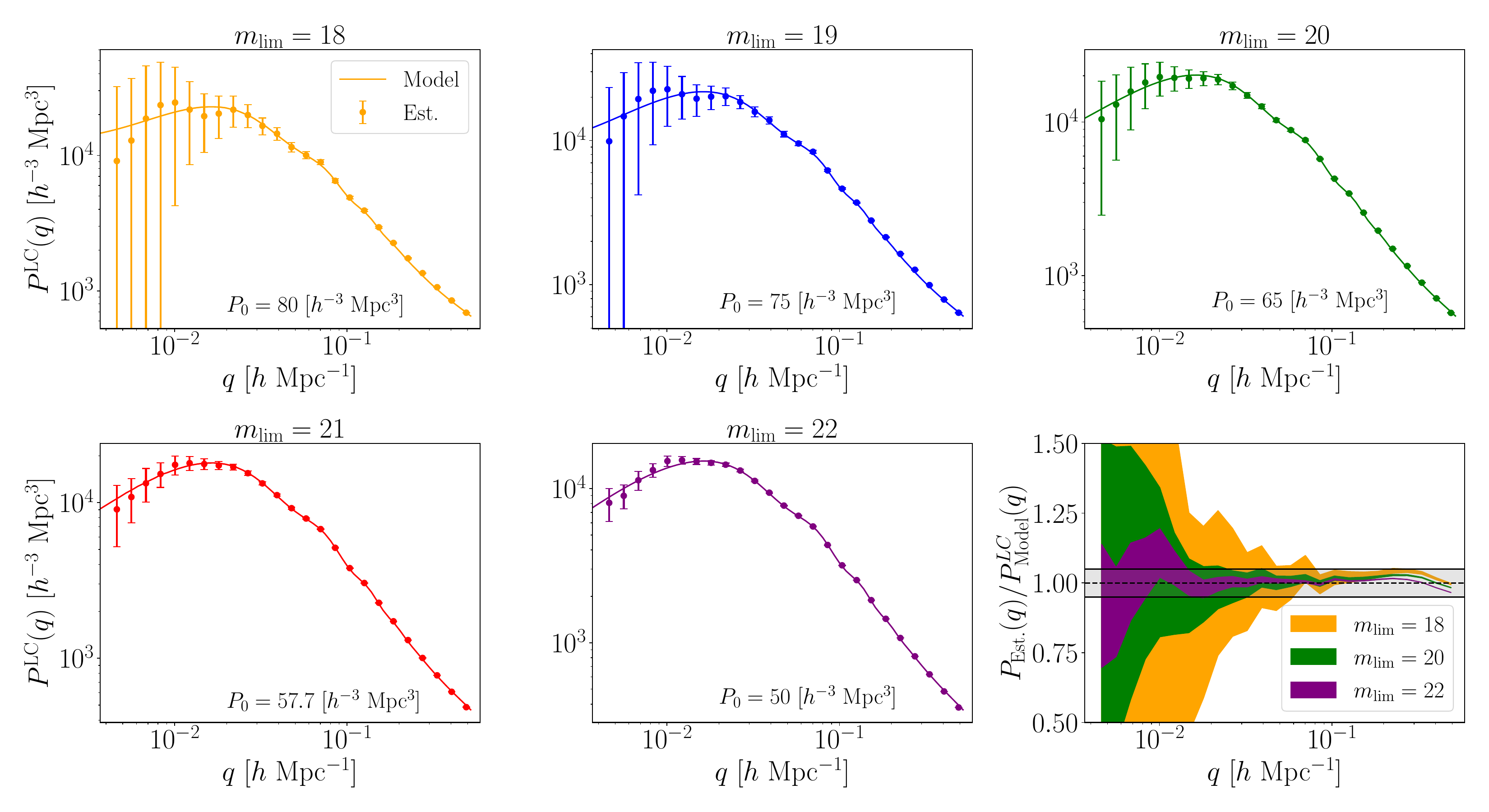} }
\caption{\small{Upper and left: Plots of the FKP-weighted power
    spectrum measurements (points with error bars) from the 5
    different magnitude cut catalogues, compared with the predictions
    from the theory (solid line). Bottom right: Ratio plots comparing
    the measurements to the model for 3 of the catalogues, with 5\%
    difference in the shaded grey region.}
\label{fig:lightcone1:pk_mags_plots_weights}}
\end{figure*}


\begin{figure}
\centerline{ \includegraphics[width=9cm]{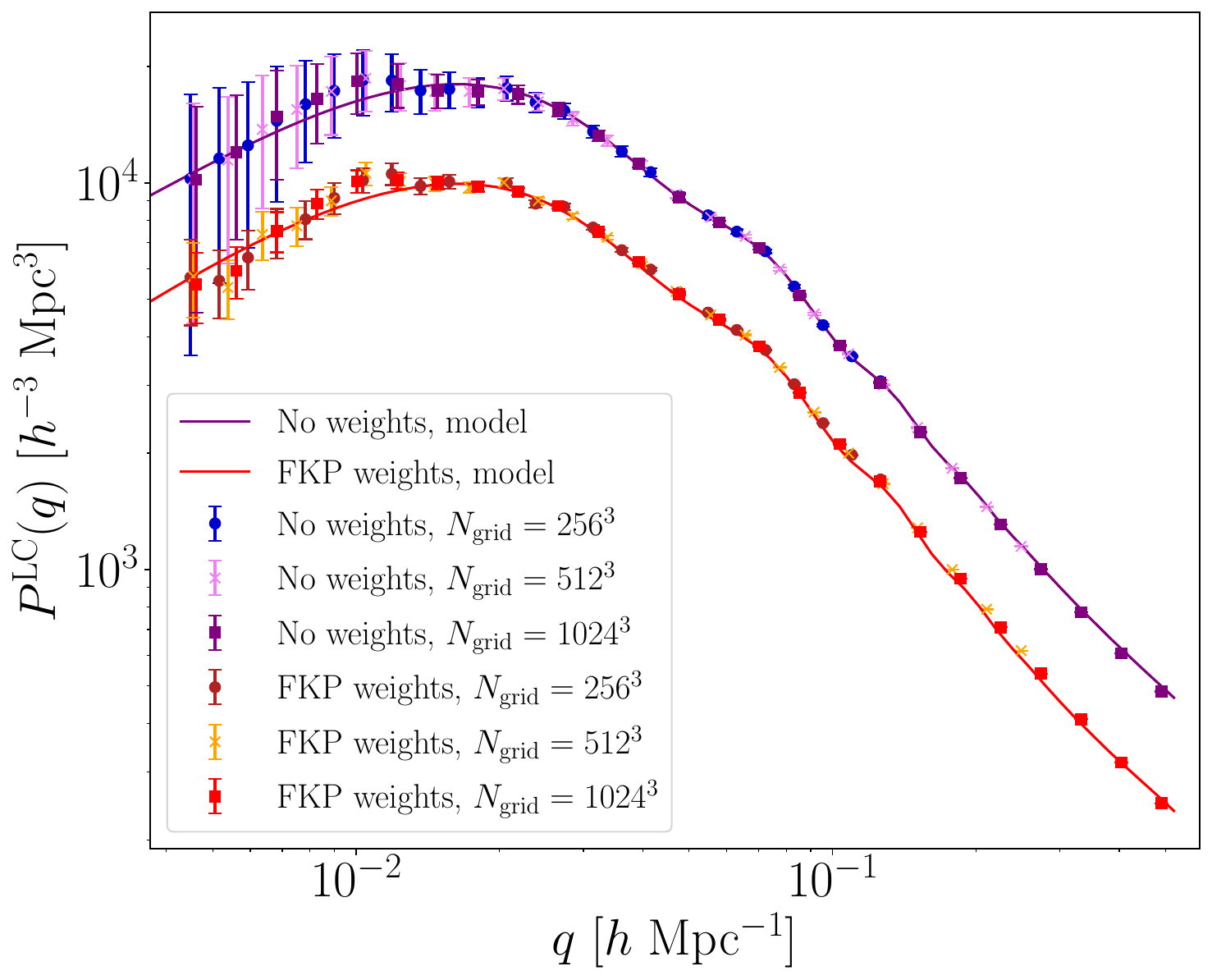} }
\caption{\small{Plot showing the effect of varying the grid resolution
    on the power spectrum measurements. We show estimates from using
    $256^3$, $512^3$ and $1024^3$ FFT grids, for both unweighted (blue
    hues) and FKP weighted (red hues) data, using the $m_{\rm lim}=22$
    catalogue, as well as the theoretical predictions (solid
    lines). Aside from a different available $k$ range of
    measurements, we see that grid resolution has little effect on the
    resulting measured power. Note that weighed measurements and model
    have been offset by a factor of $0.7$ in the vertical direction
    for clarity.}
\label{fig:lightcone1:grid_res_test}}
\end{figure}


\Figure{fig:lightcone1:pk_mags_plots_weights} shows the galaxy power
spectra measurements, averaged over the 8 realisations for each of the
5 magnitude cuts. The error bars are computed from \Eqn{eq:FKPCov2}
and the solid line shows the theoretical prediction using the recipe
described in \S\ref{sec:LC1:LCtheory}. For this set of measurements we
have employed an FKP weighting scheme where the choice of $P_0$ is
indicated in the text box on each plot, and is chosen to match the
approximate magnitude of the overall shot-noise contribution.  We
found that setting $P_0\gtrsim 10^3 h^{-3}\, {\rm Mpc}^3$ introduced
some bias to the shape of the power spectrum estimator at both the
small and large scales. We believe that this may be caused by the
optimal weights being applied as a grid based operation and not at the
particle level at the point of applying the particles to the mesh.
However, we note that \cite{Sutherlandetal1999} showed that variations
in $P_0$, and therefore the weighting functions, would lead to a shape
dependence of the power spectrum on the largest of scales. We reserve
further investigation of this issue for future work.

The bottom right subplot of the figure shows the ratio between the
measurements and the theory. The shaded regions denote the 1-$\sigma$
error envelopes for the $m_{\mathrm{lim}} = \{18, 20, 22\}$
catalogues, respectively. In this plot we also show the $\pm 5\%$
difference between theory and measurement as the shaded grey
horizontal region. For all of the magnitude cuts, we see that the
theory lies well within the error bars of the measurements at all
scales, right up to the Nyquist frequency of the grid. We do see some
slight fall off of the estimated power for scales of around $k \geq
0.2 \; [h \; \mathrm{Mpc}^{-1}]$. This is most likely due to
inaccuracies of our modelling of the 2-loop corrections to the theory
as detailed in \S\ref{ss:nonlin}.


\subsection{FFT grid resolution tests}

Owing to the fact that our full-sky lightcones are non-periodic
functions, it is interesting to examine whether there is any impact of
grid-resolution, and thus, aliasing effects on the estimated power
spectra. We examine this by focusing our attention on the deepest
$m_{\rm lim}=22$ galaxy catalogue and fixing the weighting scheme and
varying the FFT grid dimension.

\Figure{fig:lightcone1:grid_res_test} shows the results of this
exercise for the three grid resolutions $N_{\rm grid}=256^3$, $512^3$
and $1024^3$ and for the case of equal weights and FKP weights, with
$P_0 = 100 h^{-3} \rm{Mpc}^3$. We see that the estimates for the
different grid resolutions are in excellent agreement, showing a close
grouping for both the weighted and unweighted cases. We also show
again our theoretical predictions from \S\ref{ss:PkLC} for the case of
the equal weight and FKP weighting. These results indicate that the
small-scale discrepancies, noted earlier, are most likely caused by
issues with the nonlinear modelling, as opposed to any effects from
aliasing or the interlacing with the TSC charge assignment. 


\section{The turnover scale as a probe of the
  epoch of matter-radiation equality}\label{sec:turnover}

\subsection{Modelling the turnover scale}

As an application of our formalism and mock galaxy catalogue
development, we turn to the question of detecting the turnover scale
of the matter power spectrum. One of the exciting aspects of being
able to measure the power spectrum on very large scales is that it
presents us with the opportunity to probe the epoch of
matter-radiation equality in the early universe. The imprint of this
epoch can be directly linked to the peak of the power spectrum: the
theory of inflation predicts that the primordial matter power spectrum
should follow a simple power-law of the form, $P_{\rm primordial}(k)
\propto k^{n_{\rm s}}$ on very large scales. However, as the Universe
expands from a hot-dense state, fluctuation growth inside the horizon
is suppressed by radiation pressure support and radiation dominated
expansion -- this effect is commonly termed the `M\'ez\'aros Effect'
\cite[for a detailed treatment see][]{Dodelson2003}. The consequence
of this is that the peak point of the power spectrum, located at
$k_0$, corresponds to the size of the cosmological horizon at the
epoch of matter-radiation equality, and its measurement thus allows us
to constrain the redshift of this epoch. This in turn is sensitive to
both the matter density and Hubble parameters through the combination
$\Omega_m h^2$, amongst other parameters.

To explore the detectability of this scale we broadly follow the work
of \cite{BlakeBridle2005}. This was subsequently refined by
\cite{Pooleetal2013}, who applied the process to the WiggleZ Dark
Energy Survey to measure the parameters $\{k_0, z_{\mathrm{eq}},
\Omega_m h^2\}$. We refer the reader to these papers for full
details. Here we model the power spectrum at its peak as two piecewise
power laws of the form:
\begin{align}
\log_{10}P(k) & =
  \begin{cases}
            \log_{10} \big[P_0(1-\alpha x^2)\big] & \text{if $k_0 > k$}, \\
            \log_{10} \big[P_0 (1-\beta x^2) \big] & \text{if $k_0 \leq k < k_{\rm max}$},
  \end{cases}
  \label{eq:turnovermodel}
\end{align}
where 
\ba
x \equiv \frac{\ln k - \ln k_0}{\ln k_0} ,
\ea
$P_0$ is the power spectrum amplitude at turnover, and $\alpha$ and
$\beta$ are parameters that define the slope of the power spectrum on
either side of the turnover\footnote{We choose to model the power in
logarithmically spaced $k$-bins since this is a more slowly varying
function than the linear spaced bins basis. This will avoid any rapid
changes in values for the MCMC walkers.} \cite[for an alternate
  approach that employs a logarithmic power-law
  see][]{Cunnington2022}.


\subsection{Parameter estimation and feature detectability}

We now examine how well one could detect the turnover scale $k_0$ from
flux-limited full-sky, galaxy redshift surveys. We fit the above model
to three different sets of power spectra data:
\begin{itemize}
\item {\bf Case 1}: $P(k)$ measured from magnitude limited mocks, with
  equal weights (see \S\ref{sec:LC1:mocks}).
\item {\bf Case 2}: same as case 1, but where each galaxy is weighted
  by the FKP weight.
\item {\bf Case 3}: $P(k)$ are the theoretical predictions (given by
  the solid lines in \Fig{fig:lightcone1:pk_mags_plots_weights}). In
  this case the error bars are estimated from the actual
  measurements. We use this latter case to better assess biases and
  the impact of cosmic variance on the estimates.
\end{itemize}
The maximum {\em a posteriori} parameters for the turnover model of
\Eqn{eq:turnovermodel} are obtained through use of the MCMC
algorithm. To do this we make use of the Python package {\tt emcee}
\cite{Foremanetal2013}, which implements an affine-invariant ensemble
MCMC sampler. We use 1000 walkers, with an initial `burn-in' stage of
1000 steps, followed by $10^5$ iterations per walker to explore the
parameter space giving $10^8$ total propositions. We also adopt a
Gaussian likelihood model and flat uninformative priors and Table
\ref{tbl:MCMCparams} lists the prior ranges for the 4 fitted
parameters. The starting points for our MCMC walkers are distributed
in a uniform random way throughout the prior volume.  To remove the
complexity of modelling non-linear clustering effects and BAO
features, we restrict the $k$-domain of the fit to $k < k_{\rm{max}} =
0.05 \kMpc$.


\begin{table}
    \centering
    \begin{tabular}{ | c || c | c |  }
    \hline
    Parameter  &  Min  &  Max \\
    \hline
    $\log_{10}[P_0/h^{-3}{\rm Mpc^3}]$  &  3  &  5 \\
    $k_0/\kMpc$  &  $5\times10^{-3}$  &  0.05 \\
    $\alpha$  &  -5  &  10 \\
    $\beta$   &  -5  &  10 \\
    \hline
    \end{tabular}
\caption{Turnover scale model parameters and their (flat) prior
  ranges.  Column 1: parameter; columns 2 and 3 give the minimum and
  maximum posterior values. }
\label{tbl:MCMCparams}
\end{table}


\begin{figure}
\centerline{ \includegraphics[width=10cm]{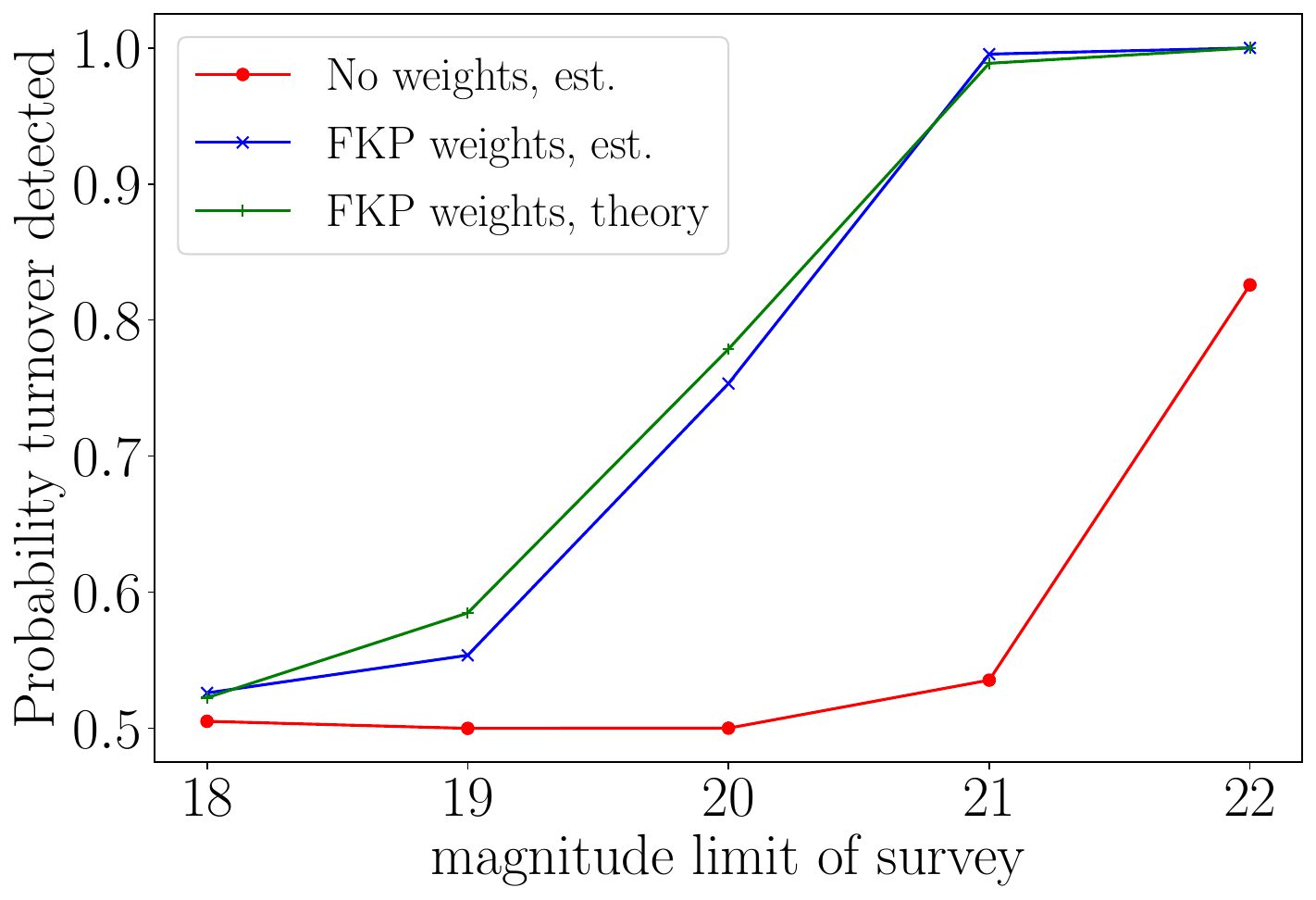}}
\caption{\small {The probability that a turnover has been measured as
    a function of the apparent magnitude limit of the survey. The case
    of no weights is indicated by the red solid points. The case of
    FKP weights is denoted by the blue crosses. The case where the
    mean measurement is replaced by the theoretical model, but again
    with FKP weighting, is given by the rotated green
    crosses. } \label{fig:lightcone1:turnover_prob}}
\end{figure}


To calculate the probability that a power spectrum turnover has been
measured, we perform a likelihood ratio test. We do this for the above
turnover model, as well as for the model where $\alpha=0$, which
signifies the `no turnover' case. For our adopted Gaussian likelihood
models, this ratio can be written as:
\be
   {\mathcal R}_{\rm rel} = \frac{L_{\rm no-turn}}{L_{\rm turn}}
   = \exp[-(\chi^2_{\rm no-turn} - \chi^2_{\rm turn})/2]\ , 
\ee
where $\chi^2_{\rm no-turn}$ and $ \chi^2_{\rm turn}$ are the
chi-square distributions:
\begin{align}
  \chi^2_{\rm no-turn} & = \sum_{i}\frac{\left[P_i-P_{\rm no-turn}(k_i)\right]^2}{\sigma^2_i}\ ; \\
  \chi^2_{\rm turn} & = \sum_{i}\frac{\left[P_i-P_{\rm turn}(k_i)\right]^2}{\sigma^2_i}\ .
\end{align}
From this information we can now derive the probability that we have
detected a turnover\footnote{We note that, strictly speaking, a
calculation of the likelihoods would require evaluation of the full
covariance matrix, however for our two large catalogues that we are
running the current analysis on, the window function is significantly
compact at and around the turnover scale as to minimise the effect of
our diagonal covariance assumption in \Eqn{eq:FKPCov2}.}. From the
constraint that $p_{\rm no-turn} + p_{\rm turn} = 1$, and that the
ratio of likelihoods is proportional to the probability ratios $L_{\rm
  no-turn}/L_{\rm turn}= p_{\rm no-turn}/p_{\rm turn}$, we have:
\ba
p_{\rm turn} = 1 - p_{\rm no-turn} = \frac{1}{1+{\mathcal R}_{\rm rel} }\ .
\ea
Thus, for each power spectrum measurement, we can calculate the
$\chi^2$ value for both models, the relative probability, and then
$p_{\rm turn} $.


\begin{figure}
\centerline{ \includegraphics[width=10cm]{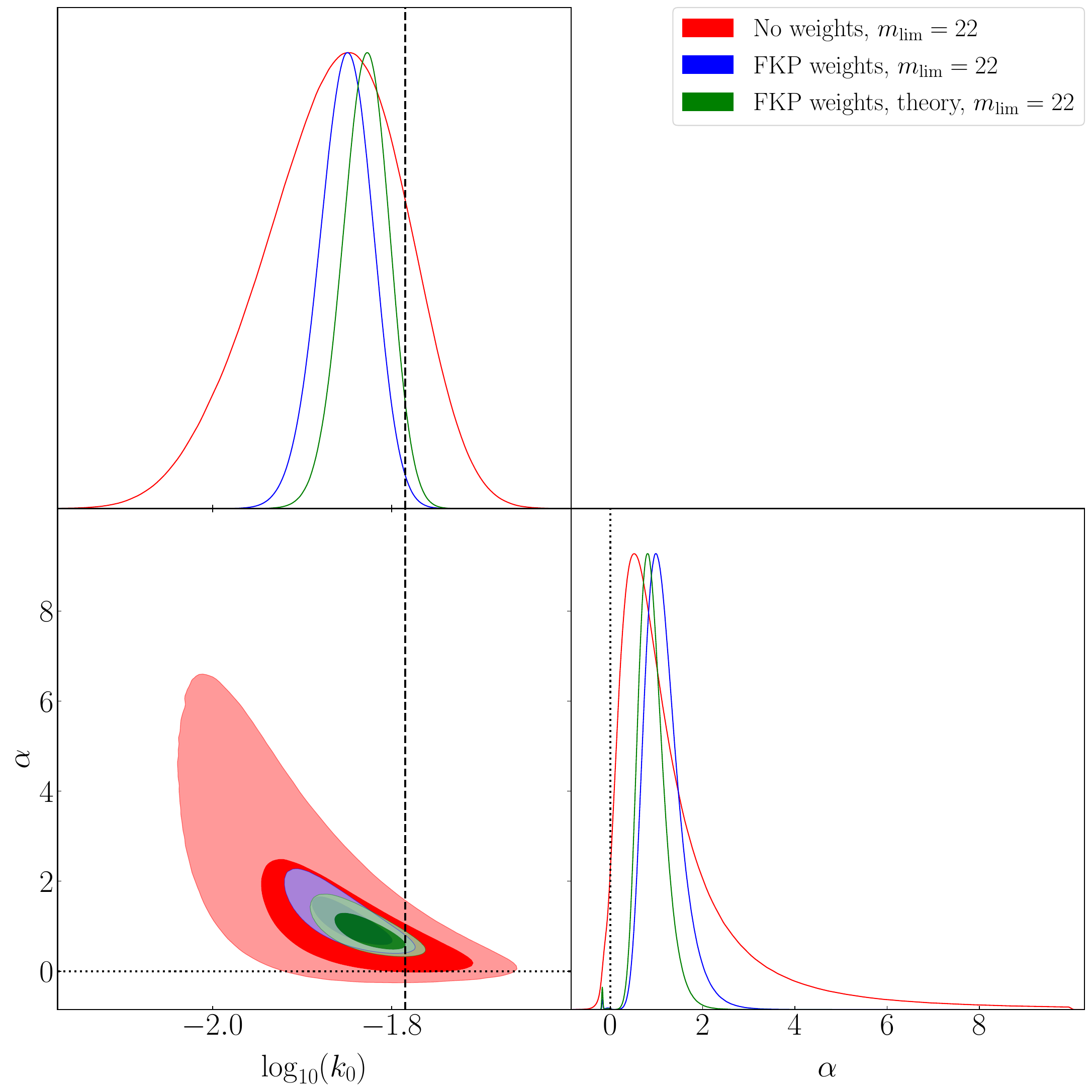}}
\caption{\small{The 1-D and 2-D marginalised posterior distribution
    for the turnover scale $k_0$ and the large-scale slope $\alpha$,
    for the $m_{\rm lim}<22$, full-sky, mock galaxy catalogues. The
    red, blue and green lines in the top left and bottom right denote
    the full 1-D posterior distributions and the shaded contours in
    the bottom left panel denote the $68\%$ and $95\%$ confidence
    regions, for the case of, no weights, FKP weights and the
    theoretical predictions with FKP weights, respectively. The dashed
    line denotes the true peak of the linear power spectrum and the
    dotted line the case of no turnover.
  } \label{fig:lightcone1:turnover_params}}
\end{figure}


\begin{figure}
\centerline{
  \includegraphics[width=11cm]{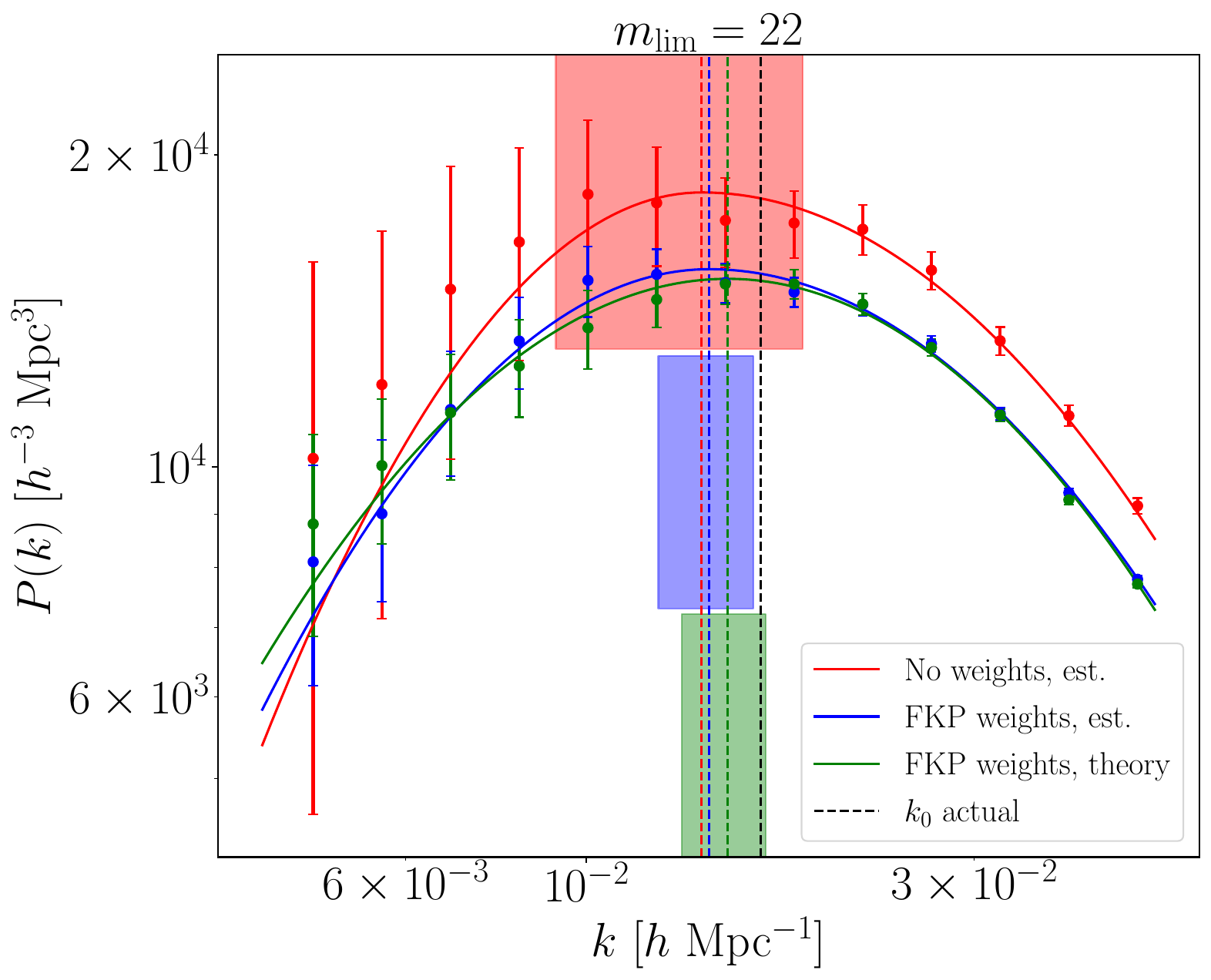}}
\caption{\small {Power spectrum as a function of wavenumber for the
    deep, full-sky $m_{\rm lim}<22$ galaxy catalogue. The data points
    show show again the measurements from the 8 mocks, but where the
    red, blue and green colours denote the cases of no weighting, FKP
    weights and FKP weights, but with the measured signal swapped for
    our theoretical predictions, respectively.  The solid lines show
    the results for the turnover model evaluated with the best-fit
    parameters. The dashed vertical lines give the the derived values
    of $k_0$ with shaded regions (with separate y ranges for clarity)
    showing the errors.  The actual value of $k_0$ obtained from the
    linear theory spectrum is shown as the dashed vertical black
    line.} \label{fig:lightcone1:turnover_best_fit}}
\end{figure}


\begin{figure}
\centerline{ \includegraphics[width=10cm]{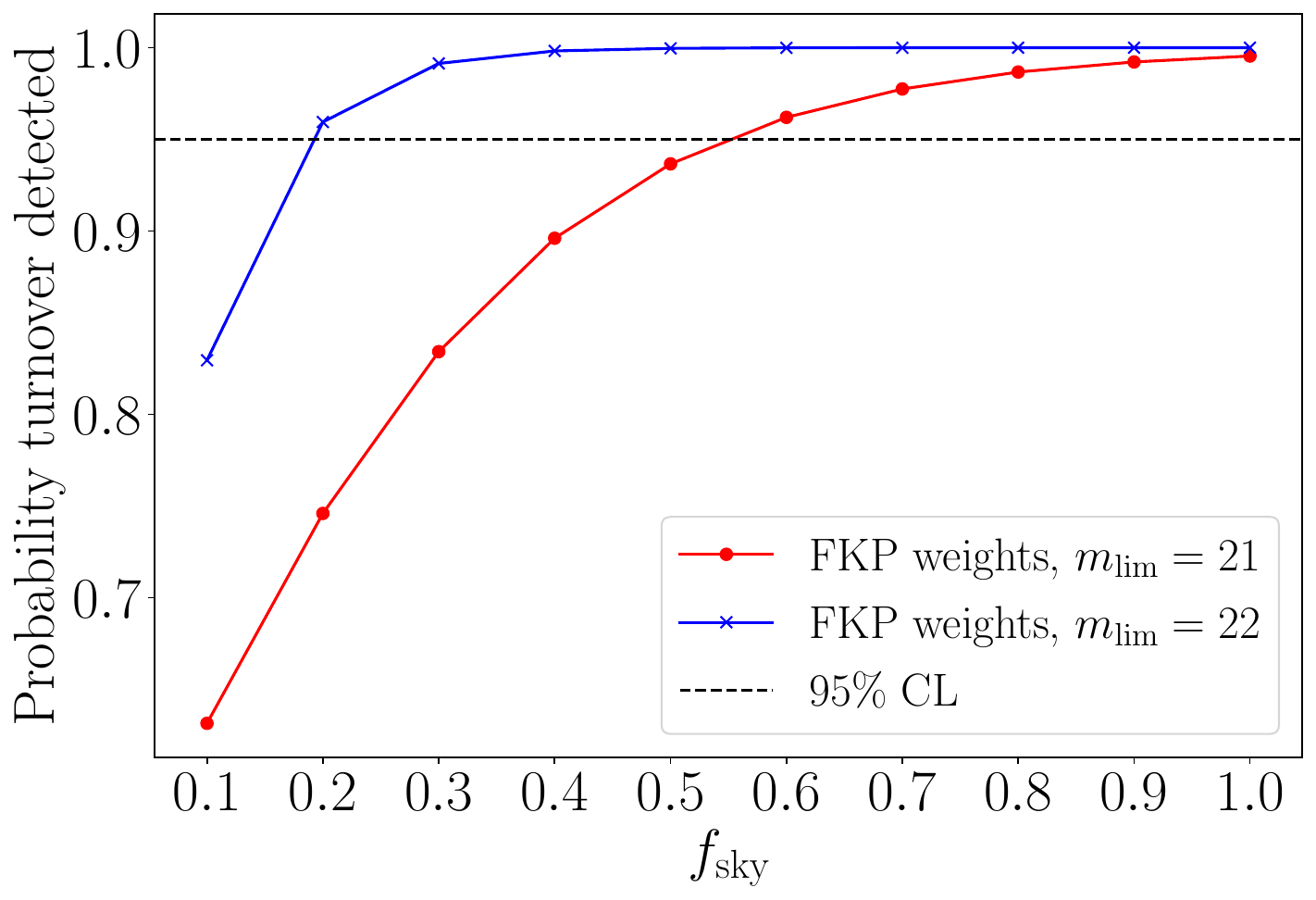}}
\caption{\small {Detection probability of the turnover scale as a
    function of the sky fraction $f_{\rm sky}$ covered by our mock
    survey data. Red and blue points and lines show the results for
    the $m_{\rm lim}<21$ and $m_{\rm lim}<22$ catalogues. The dashed
    line indicates the 95\% confidence limit
    threshold.} \label{fig:lightcone1:turnover_prob_fsky}}
\end{figure}


\Figure{fig:lightcone1:turnover_prob} shows the turnover probability
as a function of survey limiting magnitude, and for the three
different cases listed above. We see that the ability to detect a
turnover scale with 95\% confidence in our magnitude limited
catalogues is only achievable for our deepest two mocks ($m_{\rm
  lim}=21$ and $m_{\rm lim}=22$), and only when we employ the FKP
weighting scheme. This result is independent of whether we use our
actual measurements from the lightcones, or from the theoretical
predictions. Conversely, in the absence of an optimal weighting
scheme, we find that this simpler approach cannot unambiguously
discern the turnover scale.

\Figure{fig:lightcone1:turnover_params} shows the joint 1-D and 2-D
posterior distributions marginalised over all other parameters, for
the $k_0$ and $\alpha$ from the MCMC analysis, and for the deep
$m_{\rm lim}=22$ catalogue. These are the key parameters that
determine whether the data has a turnover or not.  We see that the
joint constraints from all three scenarios considered find similar
values for the turnover scale, with the FKP weighted data giving $k_0
= 0.0141 \pm 0.0015 \kMpc$.  This value is within $1.5\sigma$ of the
actual value for the turnover of the underlying linear power spectrum
$k=0.0164\kMpc$ (represented in the plot as the vertical dashed black
line). We also find that the joint constraints for the best-fit
turnover slope parameter are, for the FKP weighted data, $\alpha =
1.15^{0.25}_{-0.46}$. Thus the no turnover case corresponding to
$\alpha = 0$ can be convincingly rejected in our mock samples.

It is interesting to note that the slight bump in the measured power
spectra at around $k \simeq 10^{-2}\kMpc$, which can be seen in
\Fig{fig:lightcone1:pk_mags_plots_weights}, slightly flattens the top
of the power spectrum. This has the effect of dragging the best fit
turnover scale towards slightly larger scales and our range of values
for $k_0$ lie slightly below the actual value of $k=0.0164\kMpc$. This
analysis demonstrates some the difficulties in accurately measuring
the turnover scale \cite[for another recent look at this issue, in the
  context of HI surveys, see][]{Cunnington2022}.
    
In \Figure{fig:lightcone1:turnover_best_fit} we present again the
measured power spectra for the $m_{\rm lim}<22$ catalogues, as a
function of wavenumber. However, we now show the best-fit turnover
models, overlay the best-fit values for the turnover scale $k_0$ for
each data set and indicates the true turnover scale as measured from
the input linear theory spectrum. One can see that while the models
all fit the data very well, there is, as noted above, a small
systematic underestimate of the true $k_0$.  We partly attribute this
to large-scale fluctuations in the estimated $P(k)$ points (see the
red and blue points), which drags the fit to prefer smaller values of
$k_0$. This is much less pronounced for the case where we use the
lightcone theory predictions as $P(k)$ to determine $k_0$, however the
underestimate remains.


\subsection{Scaling to incomplete sky-coverage}

Up to this point our analysis has focused on an all-sky survey, which
is of most relevance to planned missions like {\tt SPHEREx}
\cite{Doreetal2014}. However, we now attempt to relate our study to
upcoming smaller, wide-angled, deep, galaxy redshift surveys. We do
this by examining how the probability $p_{\rm turn}$ depends on
surveyed sky fraction $f_{\rm sky}$.  Recalling \Eqn{eq:FKPCov2}, we
see that the error bars for our $P(k)$ estimates scale as $\sigma(k)
\propto 1/\sqrt{N_k}$, where $N_k$ is the number of Fourier
modes. This quantity scales as $N_k\propto \Vmu$, i.e. the survey
volume, which in turn is proportional to $f_{\rm sky}$. Thus we expect
that $\sigma(k) \propto 1/\sqrt{f_{\rm sky}}$. Owing to the fact that
we were only able to robustly detect the turnover scale for our deeper
surveys, we will also restrict our attention to the $m_{\rm lim}=21$
and 22 catalogues. We now repeat the MCMC posterior estimation, but
increasing the error bars by a factor of $1/\sqrt{f_{\rm sky}}$ for
$f_{\rm sky} \in \{1.0, 0.1\}$ in $0.1$ intervals.

\Figure{fig:lightcone1:turnover_prob_fsky} shows the detection
probability for the two deep catalogues as a function of the sky
fraction $f_{\rm sky}$. We see that the $m_{\rm lim}<22$ catalogue is
able to detect the turnover scale at a $\gtrsim 95\%$ confidence limit
for sky fractions $\gtrsim 20\%$\footnote{For comparison, the DESI
footprint has a sky fraction of around $35\%$, while for the 4MOST
cosmology redshift survey its around $18\%$.}. For the shallower
$m_{\rm lim}<21$ mock catalogue, we see that the sky fraction needs to
be $\gtrsim55\%$ for a similar level of detection.

Before moving on, it is important to note that here we have assumed
that the shape of the power spectrum remains constant as the sky
fraction is reduced. While this should hold true at smaller scales,
depending on how the survey geometry changes, we may lose the ability
to probe $k$-modes at the largest scales. This would decrease the
number of data points past the turnover scale that we have available
to constrain the model, and likely increase their errors. Furthermore,
as discussed earlier in \S\ref{sec:est}, the covariance matrix is
non-diagonal due to the mask, and so true significance must be
determined from a more advanced analysis and we leave this to future
work.


\section{Conclusions and discussion} \label{sec:LC1:conclusion}


Upcoming measurements of the large-scale structure of the Universe
will probe `effective volumes' that approach that of the entire
observable Universe \cite{Doreetal2014}. This will enable us to
perform cosmological tests to an unprecedented precision.  However, in
order to extract the information from such data sets, we must ensure
that we are also able to model the observable clustering signals with
an unprecedented degree of both accuracy and precision. This work
contributes to that effort. In this paper, we have performed a
detailed study of the geometrical lightcone effect on the two-point
galaxy clustering signal in Fourier space (the power spectrum) for a
deep, flux-limited, full-sky survey.

In \S\ref{sec:LC1:theory} we presented some key background concepts,
in particular the observable density field on the past lightcone and
the spherical-Fourier-Bessel expansion of the field.

In \S\ref{sec:LC1:LCtheory} we derived an expression for the galaxy
two-point correlation function on the past-lightcone. We also
presented the theory for the observable galaxy power spectrum, in real
space, including the effects of shot-noise and luminosity dependent
galaxy bias and for the case of an evolving luminosity function. We
showed that the mean-redshift approximation biases the amplitude and
shape of the galaxy power spectrum by more than $5\%$ for our cases
considered, when compared to a model using the unequal time
correlator.

In \S\ref{sec:approx} we developed a series of approximations to the
full analytic expressions that aided evaluation. We explored how the
expressions would simplify for time separable models of the UETC. We
derived approximations in the large-scale limit, finding that the
observed power, unlike the linear theory, asymptoted to a constant
value. On small scales, we were able to show that the observed
spectrum reduced to a more tractable form. We also examined the
fixed time approximation and showed that, provided that the time was
carefully chosen, one could minimise the errors on this, but over a
restricted range of wavenumbers and that nonlinear evolution would
always break this approximation. Here we also discussed how to extend
the model into the nonlinear regime.

In \S\ref{sec:LC1:mocks} we turned to validating our model, using
$N$-body simulations. We did this by developing an algorithm to
generate full-sky dark matter lightcones by using a quadratic
interpolation scheme to interpolate the world lines of all particles
in a piecewise fashion between consecutive snapshots. We then used
this as the base from which we generated out mock galaxy samples.  We
also developed a parallel code for estimating the power spectrum from
large data sets, making use of an interlacing technique to obtain
accurate results all the way up to the Nyquist frequency. We then 
compared these measured spectra against our theoretical predictions 
in \S\ref{sec:LC1:results} and found excellent agreement
on the scales considered. However, some small discrepancies were
found on small scales, which we attributed to the breakdown of the
specific form of our nonlinear corrections.

In \S\ref{sec:turnover} we revisited the question of detecting the
turnover scale of the power spectrum as a probe for the epoch of
matter-radiation equality in the early universe.  Following the method
outlined in \cite{BlakeBridle2005} and \cite{Pooleetal2013} we used
our mock catalogues for various flux-limited surveys to measured a
peak scale $k_0$ and the power-law slope $\alpha$ of the spectrum
red-wards of the peak. We did this by performing a standard Bayesian
parameter estimation scheme, where we explored the full parameter
space using an efficient MCMC sampler. We found that only galaxy
catalogues with $m_{\rm lim} \gtrsim 21$, in conjunction with FKP
weighting scheme, had sufficient statistical power to detect the
turnover with a probability $P \gtrsim 95\%$. By rescaling the
covariance by the inverse of the sky-fraction, $1/f_{\rm sky}$, we
recomputed the posterior distributions and found that the detection
level was maintained for surveys with $m_{\rm lim}=21$ and $m_{\rm
  lim}=22$ for sky coverages greater than $50\%$ and $20\%$,
respectively.

There are several possible directions in which this work can be
extended. First, the model that we have developed is overly
simplistic, in that we worked only in real space. In reality, the
radial positions are affected by redshift space distortions.  In this
work we have only considered the expression for the real space power
spectrum monopole. Extending the derivations from section
\S\ref{sec:LC1:LCtheory} will also be important. This will also go
hand-in-hand with development for the power-spectrum multipoles
\cite[for recent developments along these lines
  see][]{Yamamotoetal2006,Scoccimarro2015,GrasshornGebhardtJeong2020,CastorinaDiDio2022,GrasshornGebhardtDore2022}.
However, in the end, these extensions are unlikely to change our
conclusions concerning the turnover scale, since it is expected that
for such large scales these distortions would only modify the
amplitude of the signal and not the shape of the peak or its
signal-to-noise. Second, we restricted our attention to full-sky
surveys only. One should take account of the angular mask. Third,
although we included a linear luminosity dependent bias and a generic
time separable model for the UETC, more realistic treatment of galaxy
bias is needed. We were unable to explore this with our simulations,
due to resolution issues and the difficulty in interpolating halo
positions between snapshots \cite[although see][for an example of how
  this could be done.]{Mersonetal2013}.

All three of these extensions make the model in \Eqn{eq:PF11} more
complicated, and thus it also becomes necessary to improve the
numerical methods used to forward model the theory. Our current method
employs a set of adaptive numerical quadrature routines to compute the
full Bessel function integrals on large scales. This brute force
approach is somewhat slow and prone to convergence issues. A more
efficient way of such integrals would be through the {\tt FFTLog}
algorithm of \cite{Hamilton2000}.  Another interesting prospect of
being able to measure the power spectrum accurately on large scales,
particularly past the turnover scale, is the increased ability to
constrain signatures of primordial non-Gaussianity. Such measurements
remain one of the most promising probes to discriminate between
inflationary models, with the most stringent constraints currently
coming from measurements of the CMB \cite{Planck2020}. However, as we
probe out to larger volumes with late-time Universe surveys,
measurements of the large-scale structures will be highly
complimentary and competitive \cite{Doreetal2014,Muelleretal2019,
  Muelleretal2021}.


\section*{Acknowledgements}

DP acknowledges support from an STFC Research Training Grant (grant
number ST/R505146/1). RES acknowledges support from the STFC (grant
number ST/P000525/1, ST/T000473/1). AE acknowledges support from the
European Research Council (grant number ERC-StG-716532-PUNCA).  This
research used matplotlib \cite{Hunter2007} for all of the plots, and
all code was run on the DiRAC@Durham facility, managed by the
Institute for Computational Cosmology on behalf of the STFC DiRAC HPC
Facility (www.dirac.ac.uk). The equipment was funded by BEIS capital
funding via STFC capital grants ST/K00042X/1, ST/P002293/1,
ST/R002371/1 and ST/S002502/1, Durham University and STFC operations
grant ST/R000832/1. DiRAC is part of the National e-Infrastructure.


\bibliographystyle{JHEP}
\bibliography{refs}

\vspace{5mm}
 

\appendix


\section{Spherical-Fourier-Bessel expansion}\label{App:SFB}

In this appendix we provide a short overview of the
Spherical-Fourier-Bessel (hereafter SFB) expansion of the overdensity
field. To begin, consider the overdensity field $\Delta(\eta,\br)$
written as a Fourier transform over our flat-space geometry:
\be \Delta(\eta,\br) = \int \frac{\dk}{(2\pi)^3} \Delta(\eta,\bk) e^{-i\bk\cdot\br} \ .
\label{eq:delta1}\ee 
The plane wave can be expanded as a sum over spherical harmonics and
spherical Bessel functions:
\be e^{i\bk\cdot\br} = 4\pi \sum_{\ell=0}^{\infty}\sum_{m=-\ell}^{\ell}
i^\ell j_{\ell}(kr) Y_{\ell m}(\hat{\bk})Y_{\ell m}^*(\hat{\br}) \label{eq:eikr}\ .\ee
On making use of this relation in \Eqn{eq:delta1} we find:
\be \Delta(\eta,\br) = \sum_{\ell, m} (-i)^\ell Y_{\ell m}^*(\hat{\br})
\int \frac{\dk}{(2\pi)^3} 4\pi 
 j_{\ell}(kr) Y_{\ell m}(\hat{\bk}) \Delta(\eta,\bk)\ , \label{eq:Dr}\ee 
where for simplicity we have compressed the double sum. Consider now
the integral factor on the right-hand-side of the above expression,
this can be split into a radial integral and a surface
integral. The surface integral takes the form:
\be \int \diff\hat{\bk}\ Y_{\ell m}(\hat{\bk})\Delta(\eta,\bk) =
(-1)^{\ell}\Delta_{\ell m}^*(\eta,k) \ ,\ee
where
\be\Delta_{\ell m}(\eta,k) = \int \diff\hat{\bk}\ Y_{\ell m}^*(\hat{\bk})\Delta(\eta,\bk).\ee
On substituting this all back into \Eqn{eq:Dr} gives:
\ba
\Delta(\eta,\br)
& = & \sum_{\ell,m}^{\ell}
Y_{\ell m}^*(\hat{\br})\Delta_{\ell m}^*(\eta,r) 
= \sum_{\ell,m}
Y_{\ell m}(\hat{\br})\Delta_{\ell m}(\eta,r) \ ,\label{eq:Dlmr**}
\ea
where the second equality follows from the reality of the density
field, i.e. $\Delta(\eta,\br)=\Delta^*(\eta,\br)$ and where in the
above we have defined the amplitudes of the spherical harmonics as:
\ba \Delta_{\ell m}(\eta,r) & \equiv & (-i)^\ell \int
\frac{\dk}{(2\pi)^3} j_{\ell}(kr) \Delta_{\ell m}(\eta,k)
\label{eq:Dlmr**_amp}\ .
\ea
In this work we will make extensive use of these relations. 


\section{Power spectrum on the past light-cone}\label{app:PkEst}


\subsection{A fiducial galaxy redshift survey}


Let us define our fiducial galaxy survey: suppose that we have
observed $N^{\rm tot}_\g$ galaxies and to the $i$th galaxy we assign a
luminosity $L_i$, redshift $z_i$ and angular position on the sky
$\bm\Omega_i=\bm\Omega(\theta_i,\phi_i)$. If we specify the background
FLRW spacetime, then we may convert the redshift into a comoving
radial geodesic distance $\chi_i=\chi(z_i)$. A galaxy's comoving
position vector on the past lightcone may now be expressed as
$\br_i=\br(\chi_i,\bm\Omega_i)$ (c.f. \Eqn{eq:chi}).

The survey mask function depends on both the position and luminosity
of galaxies, given an adopted flux limit $m_{\rm lim}$. In this work
we shall take the angular and radial parts of the survey mask function
to be separable:
\be \Theta(\br|L) = \Theta(\bm\Omega) \Theta(\chi|L) \ .
\ee
The angular part of the mask may be written most simply as:
\be \Theta(\bm\Omega) = 
\left\{
\begin{array}{cc}
1 \ ; & [{\bm\Omega} \in \{{\bm\Omega}_{\mu}\} ] \\
0 \ ; & [\rm otherwise] 
\end{array} \ ,
\right.
\ee
where $\{\bm\Omega_{\mu}\}$ is the set of angular positions that lie
inside the survey area. In this work we shall for simplicity assume
that we are dealing with a full-sky survey and so will neglect the
angular mask. The radial mask function may be written:
\be \Theta(\chi|L) = \left\{
\begin{array}{cc}
1 \ ; & [{\chi} \leq \chi_{\rm max}(L) ] \\
0 \ ; & [\rm otherwise]  
\end{array} \ ,
\right.\label{eq:select}
\ee
where $\chi_{\rm max}(L)$ is the maximum distance out to which a
galaxy of luminosity $L$ could have been detected. 

The survey volume for galaxies with luminosity $L$ is simply the
integral of the mask function over all space:
\be \Vmu(L) = \int  \Theta(\br|L)
dV,\ee
where $dV$ is the comoving volume element at position vector $\br$
(for a flat universe $dV=d^3r=\chi^2d\Omega d\chi$) . In what follows
it will be also useful to note that the relation $\chi_{\rm max}(L)$
may be inverted to obtain the minimum galaxy luminosity that could
have been detected at radial position $\chi(z)$ in the survey.  We
shall write this as:
\be 
\left[\Lmin(\br)/ h^{-2} L_{\odot}\right] =
10^{-\frac{2}{5}\left(m_{\rm lim} -25 -M_{\odot}\right)}\left[d_{\rm L}(\br)/\Mpc\right]^2\ , \label{eq:Lmin}
\ee
where $m_{\rm lim}$ is the apparent magnitude limit of the survey,
$M_{\odot}$ is the absolute magnitude of the sun, $h$ is the
dimensionless Hubble parameter and $d_{\rm L}$ is the luminosity
distance (for a flat universe $d_{\rm L}(z)=(1+z)\chi(z)$).  Thus for
any general function ${\mathcal B(\chi, L)}$, we have the useful
integral relations:
\ba
\int_0^{\infty} dL \int_{0}^{\infty}d\chi \Theta(\chi|L) \mathcal B(\chi, L) 
& = \int_0^{\infty} dL \int_{0}^{\chi_{\rm max}(L)}  d\chi \mathcal B(\chi, L) \nn \\
& = \int_{0}^{\infty}d\chi  \int_{\Lmin(\chi)}^{\infty} dL  \mathcal B(\chi, L) \ . \label{eq:ThetaInt}
\ea


\subsection{Galaxy density field on the  lightcone as a Dirac delta function expansion}


In general the spatial density distribution of galaxies, per unit
luminosity, at position $\bx(\chi,\bm\Omega)$ on the lightcone may be
written as a sum over Dirac delta functions:
\be 
n^{\rm LC}_{\g}(\br,L)=\sum_{i=1}^{N}\delta^{\rm D}(L-L_i)\delta^{\rm D}(\br-\br_i) \ .
\label{eq:galexp} 
\ee 
Let us define the galaxy field $\Fg$, which is related to the
over-density of galaxies. This can be written:
\begin{align} 
\FLC(\br) & = A\left[ n^{\rm LC}(\br)-
  \alpha \,n^{\rm LC}_{\rm s}(\br)\right] \nn \\
& =
A\int dL w(\br,L)\Theta(\br | L)
\left[n_{\g}^{\rm LC}(\br,L)-\alpha n_s^{\rm LC}(\br,L)\right] 
\label{eq:galden} \ ,
\end{align}
where $w(\br,L)$ is a weight function that may depend on a galaxies
position in the survey and also its luminosity. $n^{\rm LC}_s(\br,L)$
represents the number density of galaxies in a mock sample that has no
intrinsic spatial correlations, and whose density is $1/\alpha$ times
that of the true galaxy field at that mass. Note that whilst the field
$n^{\rm LC}_s(\br,L)$ has no intrinsic spatial correlations it does obey a
luminosity function, which is closely related to the galaxy luminosity
function of $\left<n^{\rm LC}_{\g}(\br,L)\right>$.  The choice for the
normalisation parameter $A$ will be given later.


\subsection{Statistical properties of the past-lightcone galaxy density field}

We desire to be able to perform various statistical analyses of the
field $\FLC$. To understand how we should do this, let us
compute its ensemble average value. This can be written,
\begin{align}
  \left< \FLC(\br)\right> & = A\left[ \left<n_{\g}^{\rm LC}(\br)\right>-
  \alpha \,\left<n_{\rm s}^{\rm LC}(\br)\right>\right] \nn \\
  &= A \int dL
  w(\br,L) \Theta(\br | L)
  \left[\left<n_{\g}^{\rm LC}(\eta',\br,L)\right>-\alpha
    \left<n_s^{\rm LC}(\eta',\br,L)\right>\right]\ ,
\label{eq:Fensemb}
\end{align}
where the angled brackets denote an ensemble average in the following
sense:
\be \left< B(\left\{\br_i,L_i\right\}|\eta) \right>  \equiv 
\int \dr_1 \dots \dr_N dL_1 \dots dL_N 
p^{\rm LC}(\br_1,\dots,\br_N,L_1,\dots,L_N|\eta)
B(\left\{\br_i(\eta),L_i(\eta)\right\}) \ .
\ee
In the above $p^{\rm LC}(\br_1,\dots,\br_N,L_1,\dots,L_N)$ is the
$N$-point joint probability distribution for the $N$ galaxies being
located at the set of spatial positions $\{\br_i\}$ on the lightcone
and having the set of luminosities $\{L_i\}$. Thus, the first
expectation on the right-hand side of \Eqn{eq:Fensemb} can be written
as:
\begin{align}  
\left< n_{\g}^{\rm LC}(\br,L) \right>  & =  \int \prod_{l=1}^{N} \left\{\dr_l dL_l\right\}
p^{\rm LC}(\br_1,\dots,\br_N,L_1,\dots,L_N) \sum_{i=1}^{N}\delta^{\rm D}(\br-\br_i)\delta^{\rm D}(L-L_i) \nn \\
& =  \sum_{i=1}^{N} p^{\rm LC}(\br,L) = N p^{\rm LC}(\br,L) \ .
\end{align}
On the first line we inserted the expansion of the galaxy density
field from \Eqn{eq:galexp} and to obtain the second we integrated over
the sum of Dirac delta functions. The quantity $p^{\rm LC}(\br,L)$ can
be simplified in the following way:
\be 
p^{\rm LC}(\br,L) = p^{\rm LC}(\br) p^{\rm LC}(L|\br) 
= \frac{1}{\Vu}\frac{\phi(L|\br)}{N/\Vu} = \frac{\phi(L|\br)}{N} \ ,
\ee
where the first equality uses Bayes theorem, the second equality
assumes that galaxies can be found anywhere inside the survey volume 
-- note that this is irrespective of whether or not the galaxy would
be observable. In the above $\phi(L|\br)$ is the galaxy luminosity function
(hereafter GLF), which gives us the number density of galaxies per
unit luminosity, where the dependence on $\br$ indicates that GLF may
evolve with redshift. Thus, as desired, we conclude that
\be
\left< n^{\rm LC}_{\g}(\br,L) \right>  = \phi(L|\br) \ .
\ee

Turning to the second expectation value, we note that the only
difference between $\left<n^{\rm LC}_{\g}(\br,L) \right>$ and
$\left<n^{\rm LC}_s(\br,L)\right>$ is the artificially increased space
density of galaxies and the absence of any intrinsic clustering.
Hence, we likewise conclude that $\alpha\left< n^{\rm LC}_{s}(\br,L) \right> =
\phi(L|\br)$. Putting this all together, we arrive at the result that
$\FLC$ is a mean-zero field:
\be \left<\FLC(\br)\right>=0 \ .\ee
%

\subsection{An estimator for the galaxy power spectrum on the past lightcone}

The correlation of the Fourier modes of the observed lightcone galaxy
field $\FLC$ can be written:
\begin{align} 
  \left<\FLC(\bq_1)\FLC(\bq_2)\right>
  & = \prod_{l=1}^2 \left\{A\int \dr_l \ex^{i\bq_l\cdot\br_l}\right\} 
  \left<
  \left[ n^{\rm LC}_{\g}(\br_1)-\alpha \,n^{\rm LC}_{\rm s}(\br_1)\right]
  \left[ n^{\rm LC}_{\g}(\br_2)-\alpha \,n^{\rm LC}_{\rm s}(\br_2)\right]
  \right> \nn \\
  & = \prod_{l=1}^2 \left\{A\int \dr_l \ex^{i\bq_l\cdot\br_l}
  \int dL_1 w(\br_l,L_l)\Theta(\br_l|L_l)\right\}
  \Big[
    \left<n^{\rm LC}_{\g}(\br_1,L_1)n^{\rm LC}_{\g}(\br_2,L_2)\right>\nn \\
    &     -2\alpha\left<n^{\rm LC}_{\g}(\br_1,L_1)n^{\rm LC}_{\rm s}(\br_2,L_2)\right>
    +\alpha^2\left<n^{\rm LC}_{\rm s}(\br_1,L_1)n^{\rm LC}_{\rm s}(\br_2,L_2)\right>
    \Big] \ ,
\label{eq:PkFg}
\end{align}
where in obtaining the second line we made use of \Eqn{eq:galden} and
assumed that $\left<n^{\rm LC}_{\rm s}(\br_1,L_1)n^{\rm
  LC}_{\g}(\br_2,L_2)\right> = \left<n^{\rm LC}_{\rm
  g}(\br_1,L_1)n^{\rm LC}_{\rm s}(\br_2,L_2)\right>$.  To proceed
further we need to compute the expectations $\left<n^{\rm
  LC}_{\g}(\br_1,L_1)n^{\rm LC}_{\g}(\br_2,L_2)\right>$ and to
simplify the notation we shall use the shorthand notation $\langle
n_\g n'_\g\rangle_{\rm LC} \equiv \langle n^{\rm LC}_\g(\br, L) n^{\rm
  LC}_\g(\br', L') \rangle$, and similarly for $\langle n_\g
n'_s\rangle_{\rm LC},\, \langle n_s n'_s \rangle_{\rm LC}$.  On
inserting the delta function expansion of \Eqn{eq:galexp} we get:
\be
\left<n_\g n'_\g \right>_{\rm LC} = \int \prod_{l=1}^{N} \left\{\dr_l d L_l\right\}
p^{\rm LC}(\br_1,\dots,\br_N,L_1,\dots,L_N) \sum_{i,j}^{N}\delta^{\rm D}(\br-\br_i)
\delta^{\rm D}(L-L_i)\delta^{\rm D}(\br'-\br_j) \delta^{\rm D}(L'-L_j)  \ .
\nn \ee
The double sum can be broken up into the terms where $(i\ne j)$ and
the terms where $(i=j)$, whereupon
\begin{align}
\left<n_\g n'_\g\right>_{\rm LC} & =  \int \prod_{l=1}^{N} \left\{\dr_l dL_l\right\} 
p^{\rm LC}(\br_1,\dots,\br_N,L_1,\dots,L_N) \left[\sum_{i\ne j}^{N} 
\delta^{\rm D}(\br-\br_i)\delta^{\rm D}(L-L_i)\delta^{\rm D}(\br'-\br_j)
\delta^{\rm D}(L'-L_j)\right. \nn \\ 
& \hspace{3cm}  + \left.\sum_{i=j}^{N} \delta^{\rm D}(\br-\br_i)\delta^{\rm D}(\br'-\br_i)
\delta^{\rm D}(L-L_i)\delta^{\rm D}(L'-L_i) \right] \nn \\
& =  
\sum_{i\ne j}^{N} p^{\rm LC}(\br,\br', L, L') + \sum_{i=j}^{N} p^{\rm LC}(\br, L) 
\delta^{\rm D}(\br-\br')\delta^{\rm D}(L-L') \nn \\
& =  
N(N-1) p^{\rm LC}(\br,\br', L, L') + Np^{\rm LC}(\br, L) 
\delta^{\rm D}(\br-\br')\delta^{\rm D}(L-L') \label{eq:ncnc0} \ .
\end{align}
In order to proceed further, we need to specify the joint probability
density distribution for obtaining galaxies at positions $\br$ and
$\br'$ and with masses $L$ and $L'$. This we do through the introduction 
of correlation functions:
\begin{align}
  p^{\rm LC}(\br, \br', L, L') & \equiv p^{\rm LC}(\br, L)p^{\rm LC}(\br', L')
  \left[1+\left<\Delta^{\rm LC}(\br_1|L_1)\Delta^{\rm LC}(\br_2|L_2)\right> \right] \nn \\
  & = \frac{1}{N^2} \phi(L|\br) \phi(L'|\br')
  \left[1+\left<\Delta^{\rm LC}(\br_1|L_1)\Delta^{\rm LC}(\br_2|L_2)\right>\right] \ ,
\end{align}
where $\left<\Delta^{\rm LC}(\br_1|L_1)\Delta^{\rm LC}(\br_2|L_2)\right>$ is the
two-point cross-correlation function of galaxies with luminosities $L$
and $L'$ at positions $\br$ and $\br'$ on the past lightcone. On use
of the above definition in \Eqn{eq:ncnc0}, we find
\be
\left<n_\g n'_\g \right>_{\rm LC} \approx  \frac{}{}\phi(L|\br)\phi(L'|\br')
\left[1+\left<\Delta^{\rm LC}(\br|L)\Delta^{\rm LC}(\br'|L')\right>
  \right]+\phi(L|\br)\delta^{\rm D}(\br-\br')
\delta^{\rm D}(L-L')  \ ,
\label{eq:ncnc}
\ee
where in arriving at the last equality we have assumed that $N\gg 1$
and therefore $N-1\approx N$.  We may now write down directly the
remaining expectation values that enter \Eqn{eq:PkFg}:
\ba 
\left<n_\g n'_s\right>_{\rm LC} & = & \alpha^{-1}\phi(L|\br) \phi(L'|\br') \ ; \nn \\
\langle n_s n'_s \rangle_{\rm LC} & = & \alpha^{-2} \phi(L|\br) \phi( L'|\br')
+\alpha^{-1}\phi(L|\br)\delta^{\rm D}(\br-\br')\delta^{\rm D}(L-L') \ .
\label{eq:ncns}
\ea
On inserting these into \Eqn{eq:PkFg} we find:
\begin{align} 
\left<\FLC(\bq_1)\FLC(\bq_2)\right>
&= 
\prod_{l=1}^{2} \left\{ A\int \dr_l  \ex^{i\bq_l\cdot\br_l} \int dL_l
w(\br_l,L_l)\Theta(\br_l|L_l)\right\} \Big[  \phi(L_1|\br_1)\phi(L_2|\br_2)\nn \\
& \hspace{-1cm}\times
\left<\Delta^{\rm LC}(\br_1|L_1)\Delta^{\rm LC}(\br_2|L_2)\right>
 +\phi(L_1|\br_1)\delta^{\rm D}(\br_1-\br_2)\delta^{\rm D}(L_1-L_2) \Big] \ .
\label{eq:PkFg2}
\end{align}

Let us now focus on the first term in square brackets on the
right-hand-side of \Eqn{eq:PkFg2}, and let us assume that the
luminosity dependence of the lightcone correlation function scales in
the following way\footnote{Note that this could easily be extended to
include nonlinear bias.}:
\be \left<\Delta^{\rm LC}(\br_1|L_1)\Delta^{\rm LC}(\br_2|L_2)\right> = 
b(L_1|\chi_1)b(L_2|\chi_2)
\left<\Delta^{\rm LC}(\br_1)\Delta^{\rm LC}(\br_2)\right> \ . \label{eq:linbias}\ee
Under this assumption, the Fourier mode correlation takes the form:
\cite{Percivaletal2004a}:
\begin{align} 
\left<\FLC(\bq_1)\FLC(\bq_2)\right> 
& = \prod_{l=1}^2 \left\{ A\int \dr_l  \ex^{i\bq_l\cdot\br_l}
n^{\rm LC}_{\rm Eff}(\br_l)b^{\rm LC}_{\rm Eff}(\br_l) \right\}
\left<\Delta^{\rm LC}(\br_1)\Delta^{\rm LC}(\br_2)\right>
\label{eq:PkFg3a}
\end{align}
where the correlator $\left<\Delta(\br_1)\Delta(\br_2)\right>$ is that
of the matter on the past lightcone, as given in \Eqn{eq:corr} and
where we have defined the effective galaxy bias and number density as:
\begin{align}
  b^{\rm LC}_{\rm Eff}(\chi) & \equiv
       \frac{\int^{\infty}_{\Lmin(\chi)} dL w(\br,L)\phi(L|\br)b(L|\chi)}
            {\int^{\infty}_{\Lmin(\chi)} dL w(\br,L)\phi(L|\br)} \ ;
            \label{eq:effbias}
            \\
           n^{\rm LC}_{\rm Eff}(\chi) & \equiv \int^{\infty}_{\Lmin(\chi)} dL w(\br,L)\phi(L|\br).
           \label{eq:effden}
\end{align}
Note, in the case of no luminosity dependent biasing of clustering,
such that $b(L|\chi)=1$, then $b^{\rm LC}_{\rm Eff}(\chi)=1$ and
$n^{\rm LC}_{\rm Eff}(\chi)=w(\br)n^{\rm LC}_{0}(\chi)$.

Let us now inspect the second term in square brackets in
\Eqn{eq:PkFg2}. On integrating over $\br_2$ and $L_2$, and setting
$\bq_2=-\bq_1$ this gives the shot-noise contribution
\cite{Percivaletal2004a}:
\begin{align}
P^{\rm LC}_{\rm shot} 
& = A^2 \int \dr  \int dL w^2(\br,L) \Theta(\br|L)\phi(L|\br) \nn \\
& = 4\pi A^2 \int_{0}^{\infty} d\chi \chi^2 \int_{\Lmin(\chi)}^{\infty} dL w^2(\chi,L) \phi(L|\chi)
\label{eq:PkShot}
\end{align}
where in obtaining the above result we have made use of the fact that
$\Theta^2(\br|L)=\Theta(\br|L)$ and we have set $|\br|=\chi$. If the
weight function is assumed to be independent of galaxy luminosity,
then the shot-noise term reduces to the standard expression from FKP:
\begin{align}
P^{\rm LC}_{\rm shot} 
& \xLongrightarrow[w(\br|L)=w(\br)]{}
4\pi A^2 \int_{0}^{\infty} d\chi \chi^2 w^2(\chi) n^{\rm LC}_0(\chi)
\label{eq:PkShot2}
\end{align}
where the mean number density of galaxies on the past lightcone is
given by:
\be n^{\rm LC}_0(\chi) \equiv \int_{\Lmin(\chi)}^{\infty} dL  \phi(L|\chi) \ .\ee


\section{Modelling the evolving luminosity function and mock generation} \label{ss:nrevo2}


\subsection{The evolving luminosity function} \label{App:Schech}

In this section, we detail how the galaxy number density on the
lightcone, $n_0^{\rm{LC}}(r)$, is calculated. To do this, we use the
following formula:
\ba
n_0^{\rm{LC}}(r) = \int^{M_{\rm{lim}}(r)-5\log_{10}h}_{-\infty} \phi(M, r) dM \ .
\label{eq:nrLC}
\ea
In the above expression $\phi(M, r)$ is the evolving Schechter
luminosity function, and in terms of absolute magnitudes this has the
form:
\ba
\phi(M, r) &=& \frac{2}{5} \phi^*(r) \ln(10)
10^{\frac{2}{5}(\alpha+1)[M^*(r) - M]}\exp\big(-10^{\frac{2}{5}[M^*(r) - M]}\big) \ .
\ea
When modelling the evolution, we follow \cite{Lovedayetal2012} and let
the two parameters $M^*(r)$ and $\phi^*(r)$ evolve with comoving
radial distance, but keep the faint-end slope $\alpha$ fixed.  For the
evolving parameters we use the functional forms given in
\cite{Lovedayetal2012}:
\ba
M^*(z) &=& M^*(z_0) - Q(z-z_0)\ ; \\
\phi^*(z) &=& \phi^*(0) 10^{0.4Pz} \ .
\ea
We take the 5 free parameters to be those obtained from fitting the
$r$-band measurements for the combination of blue and red galaxies in
the phase 1 GAMA data. Values are: $P = 1.6$, $Q = 0.2$, $\alpha =
-1.23$, $M^* -5\log h= -20.7$ and $\phi^* = 0.94 \times 10^{-2} h^{3}
\mathrm{Mpc}^{-3}$. We set the fiducial redshift to $z_0 = 0.1$, which
is the redshift to which the magnitudes were $K$-corrected.

The lower limit on the integral in \Eqn{eq:nrLC} is the largest
possible absolute magnitude that is detectable, given the apparent
magnitude limit $m_{\rm lim}$, at a given comoving radial distance
$r$. This has the form:
\begin{align}
M_{\rm{lim}}(r)-5\log_{10}h & = m_{\rm{lim}} - 5\log_{10}\left(\frac{d_{\rm L}(r)}{1\Mpc}\right)-25\ .
\end{align}
We use \Eqn{eq:nrLC} in various stages of this work: first, in our
theoretical modelling of the window-convolved power spectrum on the
lightcone \Eqn{eq:PF9}. Second, we use it for sampling magnitudes for
our lightcone $N$-body simulations, so as to make magnitude-cut
catalogues to mimic galaxy samples that would be observed by a real
survey.


\subsection{Sampling luminosities from the evolving Schechter Function}  \label{App:Mags}

The method that we employ to sample luminosities from a prescribed
galaxy luminosity function can be described as follows.  First, we
draw samples from a uniform random distribution $u(R)$, where
$R\in[0,1]$. Second, these samples can be translated into random
luminosities $L$, by following the probability integral transform
method discussed in \cite{Pressetal1992}. The steps for this can be
summarised: from the transformation of random variables we have
\be | u(R) dR | = | p(L|r) dL |, \ee
where $u(R)=1$ and $p(L) = \phi(L|r)/\bar{n(r)}$, with $\phi(L|r)$
being the desired GLF and $\bar{n}(r)$ the mean number density of the
survey at comoving radial distance $r$. On integrating both sides of
the above equation we find:
\be R = \int_0^R u(R') dR' = \int_{L_{\rm min}(r)}^{L} p(L|r) dL = \Phi(L|r)\ .  \ee
Provided that we can invert the cumulative distribution function
$\Phi$, we can sample luminosities through the relation:
\be L = \Phi^{-1}(R) \ . \ee
This recipe will allow us to generate a set of luminosities whose
distribution follows the form specified by $p(L|r)$. In some cases the
inversion of $\Phi$ can be done analytically. However, for more
complex functions this can most efficiently be achieved thorough the
use of a cubic spline function to build the inverse.

In this work, we want to sample from the evolving Schechter function
(discussed in the previous section) with luminosities in the range
$L\in[L_{\rm{min}},L_{\rm{max}}]$ and for numerical ease we choose
$L_{\rm{max}} = 100 L^*$. The minimum luminosity $L_{\rm{min}}(r)$ is
determined from the apparent magnitude cut and the comoving distance
$r$ at which we are observing the galaxies (c.f. \Eqn{eq:Lmin}).  For
the Schechter function GLF we can identify the function $p(L|r)$ as
(suppressing the $r$ dependence for simplicity):
\begin{align} p(L|r) & =
  \frac{x^{\alpha}\e^{-x}(L^*)^{-1}}
       {\int_{L_{\rm{min}}}^{L_{\rm{max}}}x^{\alpha}\e^{-x} dx} \ \ \ ; \ \ \ x\equiv L/L^*\ .
\end{align}
Making use of the incomplete gamma function notation, we see that the
cumulative distribution function can then be written:
\begin{align}
  \Phi(L)
   = \int_{L_{\rm min}}^{L} p(L|) dL 
   & =
  \frac{\int_{L_{\rm min}}^{L} dx x^{\alpha}\e^{-x}}
  {\int_{L_{\rm min}}^{L_{\rm max}} dx x^{\alpha}\e^{-x}} \nn \\
  & =
  \frac{\Gamma(\alpha+1,L/L^*) - \Gamma(\alpha + 1,L_{\rm max}/L^*)}
  {\Gamma(\alpha+1,L_{\rm min}/L^*) - \Gamma(\alpha + 1,L_{\rm max}/L^*)}\ .
\end{align}
We repeat this sampling for a range of redshift values, covering the
minimum and maximum redshift of our survey, and create a larger 2-d
spline of the overall relationship.

 
\label{lastpage}

\end{document}